\documentclass{article}
\usepackage[a4paper, total={6.5in, 10in}]{geometry}
\usepackage[utf8]{inputenc}
\usepackage{graphicx}
\usepackage{hyperref}
\usepackage{tikz}
\usepackage{bbding}
\usepackage{pifont,parnotes}
\usepackage{wasysym}
\usepackage{amssymb}
\usepackage{verbatim}
\usepackage{textcomp}
\usepackage{float,authblk}
\usepackage{ifthen}
\usepackage{caption}
\usepackage{subcaption}
\usepackage{amsthm}
\usepackage{bm}
\usepackage{booktabs}

\newboolean{isarxiv}
\setboolean{isarxiv}{true}

\title{How to use Machine Learning to improve the discrimination between signal and background at particle colliders}
\author[1]{X. Cid Vidal}
\author[1,2]{L. Dieste Maro\~{n}as}
\author[1]{\'{A}. Dosil Su\'{a}rez}
\affil[1]{Instituto Galego de F\'{í}sica de Altas  Enerx\'{i}as (IGFAE), Universidade de Santiago de Compostela, 15782, Santiago de Compostela (Spain)}
\affil[2]{Triple Alpha innovation, Cuns, 4, 15286, Outes. A Coru\~{n}a (Spain)}

\date{}

\makeatother

\begin{document}
\maketitle
\begin{abstract} 
 The popularity of Machine Learning (ML) has been increasing in the last decades in almost every area, being the commercial and scientific fields the most notorious ones. Concerning particle physics, ML has been proved as a useful resource to make the most of projects such as the Large Hadron Collider (LHC). The main advantage provided by ML is reducing the time and effort put into the measurements done by experiments, while improving the performance. With this work we aim to encourage scientists at particle colliders to use ML and to try the different alternatives we have available nowadays, focusing in the separation between signal and background. We assess some of the most used libraries in the field, like Toolkit for Multivariate Data Analysis with ROOT, and also newer and more sophisticated options like PyTorch and Keras. We also check how optimal are some of the most common algorithms for signal-background discrimination, such as Boosted Decision Trees, and propose the use of others, namely Neural Networks. We compare the overall performance of different algorithms and libraries in simulated LHC data and produce some guidelines to help analysts deal with different situations. Examples are the use of low or high-level features from particle detectors or the amount of statistics available for training the algorithms. Our main conclusion is that the algorithms and libraries used more frequently at LHC collaborations might not always be those providing the best results for classification of signal candidates, and fully connected Neural Networks trained with Keras can improve the performance scores in most of the cases we formulate.
  \end{abstract}
\newpage

%%%%%%%%%%%%%%%%%%%%%%%%%%%%%%%%%%%%%%%%%%%%%%%%%%%%%%%%%%%%%%%%%%%%%%%%%%%%%%%%%%%%%%%%%%%%%%%%%%%%%%%%
%%%%%%%%%%%%%%%%%%%%%%%%%%%%%%%%%%%%%%%%%%%%%%%%%%%%%%%%%%%%%%%%%%%%%%%%%%%%%%%%%%%%%%%%%%%%%%%%%%%%%%%%
%%%%%%%%%%%%%%%%%%%%%%%%%%%%%%%%%%%%%%%%%%%%%%%%%%%%%%%%%%%%%%%%%%%%%%%%%%%%%%%%%%%%%%%%%%%%%%%%%%%%%%%%
%%%%%%%%%%%%%%%%%%%%%%%%%%%%%%%%%%%%%%%%%%%%%%%%%%%%%%%%%%%%%%%%%%%%%%%%%%%%%%%%%%%%%%%%%%%%%%%%%%%%%%%%
%%%%%%%%%%%%%%%%%%%%%%%%%%%%%%%%%%%%%%%%%%%%%%%%%%%%%%%%%%%%%%%%%%%%%%%%%%%%%%%%%%%%%%%%%%%%%%%%%%%%%%%%

\newcommand{\bmu}{B\rightarrow\mu^{+}\mu^{-}}
\newcommand*{\bpipi}{B\rightarrow\pi^{+}\pi^{-}}
\newcommand*{\btripi}{B\rightarrow3\pi}
\newcommand*{\bcpi}{B\rightarrow4\pi}

\ifthenelse{\boolean{isarxiv}}
{\newcommand{\myfigwidth}{0.8}}
{\newcommand{\myfigwidth}{0.6}}

\ifthenelse{\boolean{isarxiv}}
{\newcommand{\mytabwidth}{0.8}}
{\newcommand{\mytabwidth}{0.6}}

\section{Introduction}
%\label{sec:intro}

Particle physics experiments, and especially those at particle colliders, have to deal with vast amounts of data where, very often, an elusive signal must be found against a much larger background. %please check intended meaning has been retained
This has naturally paved the way for the usage of Machine Learning (ML) in the field. As in other problems where ML applies, the separation between signal and background relies on several variables (features) that behave differently in both categories. ML algorithms are able to statistically combine the separation power of different features, making the best of them all, and using the correlations between them in its favor. The resulting discrimination power is usually superior to anything a ``manual'' selection of requirements would achieve, and can be obtained in a more efficient way. The use of ML in particle physics is an emerging area, which is extending to more and more fields, as we shall see. A very complete (and live) review of this wide range of uses can be found in Reference~\cite{Feickert:2021ajf}.

To better understand the implementation of ML as a way of improving the discrimination between signal and background, we chose the LHCb experiment \cite{LHCb:2008vvz,LHCb:2014set} at CERN as a benchmark. LHCb is one of the large collaborations in the Large Hadron Collider (LHC) project, where protons are accelerated to ultrarelativistic energies ($\sim$14 TeV) and then made to collide against each other, allowing for the study of the smallest pieces of matter. These collisions happen at an incredibly high rate ($\sim 40$ million times per second), giving rise to hundreds of other particles whose interaction is then recorded by detectors (such as LHCb), which act as gigantic photographic cameras that collect  90 petabytes of data per year~\cite{cern_data}. From these vast datasets, scientists try to extract interesting and rare phenomena, such as undiscovered particles or very infrequent particle decays, whose characteristics help to understand the properties of the smallest pieces of matter. 

The usual pipeline in analyses at a LHC experiment is as follows. Particles produced by collisions leave different types of electronic signals at the different components of the detectors (or subdetectors). These subdetectors are specialized in different aspects, from tracking to particle identification (PID). Their final goal is to provide  as realistic a picture as possible of what happened after the collision. This picture involves knowing the energy, identity and position of the particles produced, and is usually referred to as ``reconstruction''. To reconstruct each collision, different algorithms help to interpret the electronic signals grouping the particle “hits” into tracks or understanding which type of particle left those hits. Another type of object that is often reconstructed at particle colliders are jets, which are groups of particles flying very close to each other and usually originating from a single energetic particle of interest. Determining the nature of this particle from the properties of jets (known as tagging) is also a frequent task in the reconstruction of collisions. Although they are not the goal of this study, all these aspects have recently led to surge in the use of ML. Recent examples are the use of graph neural networks (aspects of some of these ML algorithms are discussed later in the paper) for PID \cite{Qasim:2019otl}, reinforcement learning for jet reconstruction \cite{Cranmer:2021gdt} or deep learning for track reconstruction~\cite{Goncharov:2021wvd} or tagging of jets \cite{Andrews:2021ejw}. We refer one more time to Reference~\cite{Feickert:2021ajf} for a more complete compendium of examples.

Once particle collisions are reconstructed, analysts are faced with a list of objects they must use to conduct measurements of interest (to perform what particle physicists denote ``analyses''). Note that this reconstruction is not perfect, but instead limited by the resolution of the detectors. Although efforts can be made to improve this resolution, such as improving the calibration of detectors by means of convolutional neural networks \cite{Akchurin:2021ahx}, the reconstructed objects will always be different to their ``original'' counterparts, which limits what experiments can do with them.  One of the main consequences of this is that, very often,  it becomes very hard to distinguish the target signals that experiments look for from backgrounds composed of other particles with similar properties. For instance, if one looks for an exotic new particle decaying into a pair of hadrons, %please check intended meaning has been retained
 this could be faked by situations in which each hadron originates from a different standard particle, although they are located close to each other in space. This all fits very nicely with the use of ML algorithms, particularly with classifiers that are able to separate scarce signals hiding behind much larger amounts of background. Finding the best of these classifiers  has been the subject of many studies in particle physics, using algorithms of very different kind, ranging from boosted decision trees \cite{Cornell:2021gut} to deep learning \cite{Baldi:2014kfa}. Alternative involves unsupervised \cite{Dillon:2020quc} or semi-supervised \cite{Dahbi:2020zjw} algorithms, with the goal of optimizing searches for rare phenomena for which  more traditional methods have failed to date. Despite the many efforts being conducted in this area of signal-background discrimination by means of ML, as we shall see, most analysts either still rely on more traditional approaches or do not use the latest, best-performing, tools available in terms of libraries and algorithms. %please check intended meaning has been retained
Assessing this question in detail is one the main goals of this~paper.

To some extent, the LHCb experiment has pioneered the application of ML at the LHC. This involves, for instance, the use of Multi-Variate Analysis (MVA) classification libraries in some of their first analyses \cite{LHCb:2011arj} or the introduction of ML in the online trigger system of the detector \cite{Williams:1323557,Likhomanenko:2015aba}, which runs automatically at every LHC proton collision.
Even if LHCb is, in many ways, an example of good use of ML in particle physics, in this paper we first assess the extent to which ML is used in their analyses and then what  kind of libraries and algorithms are currently being used in the experiment. We also compare these aspects to those of other experiments. We then contrast the performance of some of the most frequent ML libraries and those used in particle physics with simulated LHCb data, to find out the extent  to which there is room for improvement in the signal-background discrimination achieved in standard analyses. %please check intended meaningh as been retained
 Note this is a novelty, since many other comparisons of this kind rely on existing High-Energy Physics (HEP) datasets designed for benchmarking, but corresponding to other topologies, kinematic ranges and experiments \cite{higgs_dataset,Aarrestad:2021oeb,Kasieczka:2021xcg}. Next, we extend these tests to other libraries and algorithms. We quantify the overall margin of improvement and provide guidelines on how to use the latest libraries available in ML to improve the sensitivity of particle physics experiments. The conclusions of this study can mostly be expanded to experiments beyond LHCb. Therefore, instead of testing a specific algorithm to solve a specific problem, we try to provide a more global perspective, based on the actual popularity of algorithms and libraries in experiments and their potential performance. In summary, our main goals are:
\begin{itemize}
    \item Finding out the popularity of different algorithms and libraries at LHC experiments.
    \item Determining whether the most popular methods are those that provide the best performance.
    \item Providing examples of potential alternatives, which might have a better performance.
    \item Describing how these alternatives might depend on the conditions of the analysis, such as the amount of statistics that are available for training.
\end{itemize}

This article is divided as follows. In Section~\ref{sec:ML_intro} we present the main libraries and tools one can use for signal-background discrimination and check their use in different LHC experiments. Section~\ref{sec:data} introduces the datasets we simulated to perform the comparisons in the rest of the paper, as well as the main features that we used to discriminate signal against the background. Section \ref{sec:tmva_sklearn} presents a first comparison of the performance of different ML libraries for classification, out of those presented in Section~\ref{sec:ML_intro}, when facing simulated data. This comparison is extended in Section~\ref{sec:comparison} to more libraries and algorithms. We then discuss our results in Section~\ref{sec:results}. Finally, we conclude in Section~\ref{sec:conclusions}.

%%%%%%%%%%%%%%%%%%%%%%%%%%%%%%%%%%%%%%%%%%%%%%%%%%%%%%%%%%%%%%%%%%%%%%%%%%%%%%%%%%%%%%%%%%%%%%%%%%%%%%%%
%%%%%%%%%%%%%%%%%%%%%%%%%%%%%%%%%%%%%%%%%%%%%%%%%%%%%%%%%%%%%%%%%%%%%%%%%%%%%%%%%%%%%%%%%%%%%%%%%%%%%%%%
%%%%%%%%%%%%%%%%%%%%%%%%%%%%%%%%%%%%%%%%%%%%%%%%%%%%%%%%%%%%%%%%%%%%%%%%%%%%%%%%%%%%%%%%%%%%%%%%%%%%%%%%
%%%%%%%%%%%%%%%%%%%%%%%%%%%%%%%%%%%%%%%%%%%%%%%%%%%%%%%%%%%%%%%%%%%%%%%%%%%%%%%%%%%%%%%%%%%%%%%%%%%%%%%%
%%%%%%%%%%%%%%%%%%%%%%%%%%%%%%%%%%%%%%%%%%%%%%%%%%%%%%%%%%%%%%%%%%%%%%%%%%%%%%%%%%%%%%%%%%%%%%%%%%%%%%%%

\section{Main ML algorithms for signal and background discrimination and their use at particle colliders}
\label{sec:ML_intro}
When tackling signal-background discrimination by means of ML, most analysts in LHC experiments automatically  limit themselves to one library,  TMVA \cite{TMVA:1019880}, and one algorithm, BDTs with AdaBoost. In this section, we describe both of these and discuss other convenient options on the market, in terms of libraries and methods.

The Toolkit for Multivariate Data Analysis (TMVA) used with ROOT \cite{rene_brun_2019_3895860}, is an open-source data analysis framework  that provides all the necessary functionalities for
processing large volumes of data, statistic analysis, visualization and information storage. ROOT was created at CERN and, although it was designed for HEP, it is currently used in many other areas such as biology, chemistry, astronomy, etc. As ROOT is specifically designed for HEP, the library is very popular and well-known for  LHC experiments. The same is true for TVMA. As both ROOT and TMVA were developed some time ago, some of the latest techniques and innovations concerning data analysis are not yet available using just these tools. In this regard, several different TMVA interfaces have been recently created to solve this issue, with many framework options developed to integrate TMVA with more sophisticated libraries, such as Keras and PyTorch (described below). Thanks to this expansion, the use of ML for particle physicists is becoming easier, wider and more common. TMVA can be convenient for Python users, providing easy interfaces to examine data correlations, overtraining checks and a simpler event weighting management. Even though these characteristics are available in other Python libraries, TMVA offers the possibility of integrating every step of the process without the need for additional ones. Moreover, it offers suitable pre-processing possibilities for the data before feeding them into any of the classifiers. Thiese data must be given in form of either ROOT { \verb|TTrees| or ASCII text files. The first is the usual format in which scientists at CERN deal with their datasets. The analyses in this paper are based on ROOT v6.22/08.

Sklearn \cite{pedregosa2011scikit} is a native open-source library for Python, and is currently an essential tool for modern data analysis. It includes a large range of tools and algorithms, which allow for the appropriate statistical modeling of systems. It incorporates algorithms for classification, regression, clustering, and dimensionality reduction, and supports a wide variety of algorithms \cite{Geron2017Handson} such as KNNs, boosted decision trees, random forests and SVMs among others. Furthermore, it is compatible with other Python libraries such as matplotlib \cite{Hunter:2007}, pandas \cite{mckinney2010data}, SciPy \cite{2020SciPy-NMeth} or NumPy \cite{harris2020array}. %
Contrary to TMVA, Sklearn was originally designed for Python. This means that the library provides a variety of modules and algorithms that facilitate the learning and work of the data scientist in the early stages of its development.  
For TMVA users, Sklearn can provide a versatile and simple interface for ML, with very simple applications for the trained classifier in new datasets.  In this paper, we use Sklearn version 0.20.4.

Other popular types of libraries were designed to better deal with Neural Networks (NNs), which are introduced below.
PyTorch \cite{NEURIPS2019_9015} created by Adam Paszke, is object-oriented, which allows for dealing with NNs in a natural and convenient way.  An NN in PyTorch is an object, so each NN is an instance of the class through which the network is defined, all of which are inherited from the torch NNs Module. The torch NNs Module provides the main tools for creating, training, testing and deploying an NN. In our analyses, we use PyTorch 1.8.0.
 Keras \cite{chollet2015keras} is an open-source software library that provides a Python interface for artificial NNs. It is capable of running on top of Tensorflow \cite{tensorflow2015-whitepaper}, Theano \cite{2016arXiv160502688full} and CNTK \cite{10.1145/2939672.2945397}. Keras has wide compatibility between platforms for the developed models and excellent support for multiple GPUs. For this paper, we use Keras Release 2.6.0.

Moving on to the methodologies, as explained above, BDTs are among the most popular at LHC experiments. BDT stands for ``Boosted Decision Tree'', and is currently one of the most popular algorithms for classification in ML. BDTs are based on the combination of weak predictive models (decision trees or DTs) to create a stronger one. These DTs are generated sequentially, with each tree being created to correct the errors in the previous one. This process of sequencing is known as ``Boosting'', and  can be of different types. The usual trees in BDTs are typically ``shallow'', which means they have just one, two, or three levels of deepness. BDTs were designed to improve the disadvantages of the regular DTs, such as as a tendency to overfit and their sensitivity to unbalanced training data. %please check intended meaning has been retained
When boosting DTs, a very common choice is Adaptive Boosting (AdaBoost). This boosting technique is one of the canonical examples of ensemble learning in ML. In this method, the weights that are assigned to the classified instances are re-assigned at each step of the process.  AdaBoost gives higher values to the more incorrectly classified instances in order to reduce bias and the variance. With this, one can can make weak learners, such as DTs, stronger.
The main difference between other types of boosting and AdaBoost relates to the use of particular loss functions. BDTs boosted with AdaBoost are the most frequent algorithm used at the LHC for signal-background discrimination by means of ML. Due to this, in the first part of the paper, we focus on how the performance obtained by this algorithm depend on the library used (namely, TMVA vs.~Sklearn).

A very popular alternative in ML is  NNs. This is a type of model that can try to emulate the behaviour of the human brain, whose potential for HEP was first introduced decades ago \cite{DENBY1988429}.
In NNs, nodes, usually denominated ``artificial neurons'', are connected to each other to transmit signals. %
The main goal is to transmit these signals from the input to the end in order to generate an output.  The neurons of a network can be arranged in layers to form the NN. Each of the neurons in the network is given a weight, which modifies the received input. This weight changes the values that come through, to later continue on their way through the network. Once the end of the network has been reached, the final prediction calculated by the network is shown as the output. In general, it can be said that an NN is more complex the more layers and neurons it has. 

Although they are not directly part of this study, for reference, we now briefly review some of the algorithms that data scientists have developed more recently. Even though some of these are slowly being incorporated into HEP, they are beyond the scope of our analysis. One example of these new developments is that of
Extreme Gradient Boosting (XGBoost) \cite{xgboost}, which is yet another type of boosting. XGBoost modifies the more traditional BDT algorithms by using a different  objective function, so that a convex loss function is combined with a penalty term that accounts for the model complexity. XGBoost can be used together with different types of techniques, such as independent component analysis, gray wolf optimization or the whale optimization algorithm, to improve the general BDT results \cite{icax,woa}.  
Going beyond this, we should now mention ``liquid learning''. Liquid Learning \cite{Hasani_Lechner_Amini_Rus_Grosu_2021} is a new type of ML algorithm that continuously adjusts depending on new data inputs. This means that the algorithm is able to modify its internal equations do that it can always adapt to changes in the incoming data stream. The fluidity of this new type of learning makes the algorithms more robust against noise or unforeseen changes in the data. 
Finally, one of the most expected developments these days concerns quantum technologies, and, with these, quantum ML (QML) \cite{qml}. QML algorithms and models try to use the advantages of quantum technologies to improve classical ML, for instance, by developing efficient implementations of slow classical algorithms using quantum computing.

In the process of analyzing how to make the most of ML in analyses at particle colliders, as mentioned above, we first compare the two best-known libraries in particle physics, TMVA and Sklearn, to quantify which one is more effective for the same problem, i.e., using the same data with the same characteristics and the same algorithm: a BDT with AdaBoost. %please check intended meaning has been retained
 For the second part of the paper, we increase the range of methodologies beyond the usual BDTs with AdaBoost to include NNs, and enlarge the range of libraries to PyTorch and Keras. The goal is to determine how all of these methods and libraries perform in terms of signal-background discrimination. 

To compare the performance of different classifiers, the usual methodology in ML involves the use of the so-called ROC curve.
A Receiver Operating Characteristic (ROC) curve is a representation of a classifying model, built to show the discrimination power of the classifier at different thresholds. The two parameters represented in the curve are the %
rate of true positives vs.~the rate of false positives. We built ROC curves for all the classifiers of interest in this paper. For each classifier, one can also integrate the area under the corresponding ROC curve (AUC). The AUC turns out to be a very useful benchmark when one wants to compare the performance of different classifiers, providing a metric to quantify the separation power between different classes. In general, the higher the AUC score, the better the model, i.e., the more often it correctly assigns each instance to the class to which it belongs. 
This parameter is taken into account to compare the different classifiers. Moreover, we also account for the learning time needed for the algorithms to process the data, as well as the correlations with some key independent or ``spectator'' variables, such as the invariant mass.

Before moving to the problem of creating the models, let us examine how different LHC experiments have been using ML in recent years. This is a counting exercise, with which we intend to evaluate the popularity of different ML algorithms and libraries, without directly comparing their performance, which will be performed in the following sections. After looking into all LHCb publications from 2010 to August 2021, we calculated the percentage of papers per year that use TMVA and the percentage of papers that use Sklearn or other popular libraries that deal with NNs, such as Keras and PyTorch. We also generated a label, which we named ``Generic NN'' for those papers in which the use of a Neural Network is implied but the library used is not mentioned. The result can be seen in Figure~\ref{fig:lhcb_p}, which shows how the use of TMVA is higher than Sklearn.
%There is also a modest increase in the use of ML libraries in the last years, but still it is clear that not a majority of LHCb papers rely on ML.%
In Figures~\ref{fig:atlas_p} and \ref{fig:cms_p}, we repeated this for ATLAS \cite{ATLAS} and CMS \cite{CMS}, respectively. We observed a similar behaviour to LHCb, where the use of TMVA was preferred to the other libraries. In ATLAS, even though TMVA is still the most used tool, we can appreciate an increase in the use of Keras in the last two years. Note than, in all three experiments, the use of Keras, PyTorch or other types of general NNs usually was not focused on direct signal-background discrimination at the analysis level, but instead on other tasks related to the reconstruction of events at the LHC, such as jet reconstruction and tagging, PID or track reconstruction.
%\vspace{-18pt}
\begin{figure}[H]
   % \centering 
    \ifthenelse{\boolean{isarxiv}}{
    \includegraphics[width =  \textwidth]{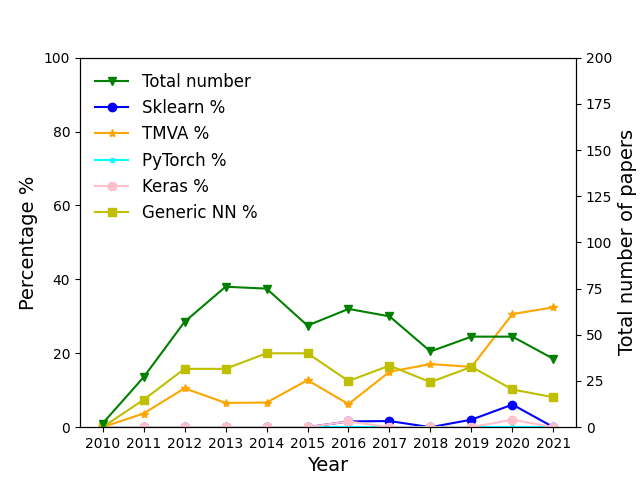}}{
    \includegraphics[width =  0.7\textwidth]{images/lhcb_2.png}}
    \caption{Usage of ML at LHCb across the years. We show  the number of papers published every year (denoted as ``Total number'' and corresponding to the right axis), as well as the fraction of them reporting the use of TMVA, Sklearn, Keras, PyTorch and Generic NNs (in the left axis). The latter category corresponds to papers mentioning the use of NNs but never referencing any of the aforementioned libraries.}
    \label{fig:lhcb_p}

\end{figure}
\vspace{-7pt}
\begin{figure}[H]
   % \centering
    \ifthenelse{\boolean{isarxiv}}{\includegraphics[width = \textwidth]{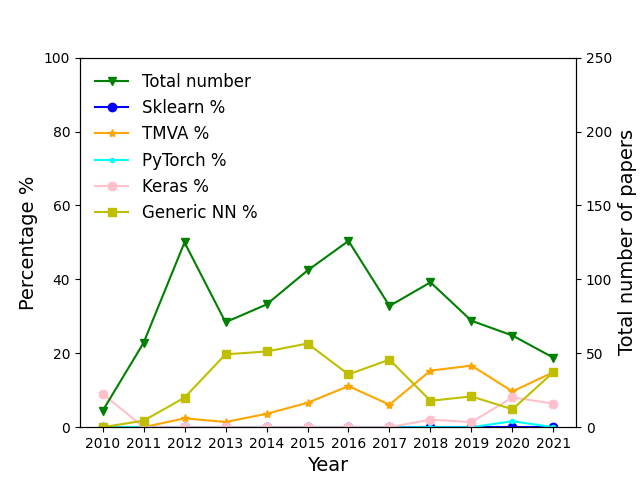}}{\includegraphics[width = 0.7\textwidth]{images/atlas_2.png}}
    \caption{Usage of ML at ATLAS across the years. We show  the number of papers published every year (denoted as ``Total number'' and corresponding to the right axis), as well as the fraction of them reporting the use of TMVA, Sklearn, Keras, PyTorch and Generic NNs (in the left axis). The latter category corresponds to papers mentioning the use of NNs but never referencing any of the aforementioned libraries.
    \label{fig:atlas_p}}
\end{figure}

\begin{figure}[H]
   % \centering
    \ifthenelse{\boolean{isarxiv}}
    {\includegraphics[width = \textwidth]{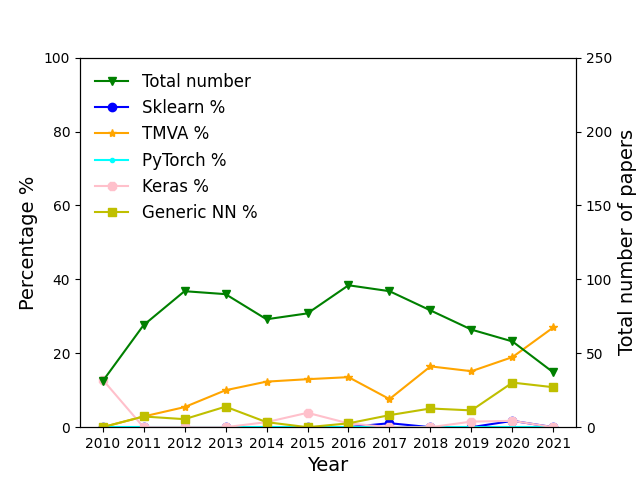}}
    {\includegraphics[width = 0.7\textwidth]{images/cms_2.png}}
    \caption{Usage of ML at CMS across the years. We show  the number of papers published every year (denoted as ``Total number'' and corresponding to the right axis), as well as the fraction of them reporting the use of TMVA, Sklearn, Keras, PyTorch and Generic NNs (in the left axis). The latter category corresponds to papers mentioning the use of NNs but never referencing any of the aforementioned libraries.}
    \label{fig:cms_p}
\end{figure}

%%%%%%%%%%%%%%%%%%%%%%%%%%%%%%%%%%%%%%%%%%%%%%%%%%%%%%%%%%%%%%%%%%%%%%%%%%%%%%%%%%%%%%%%%%%%%%%%%%%%%%%%
%%%%%%%%%%%%%%%%%%%%%%%%%%%%%%%%%%%%%%%%%%%%%%%%%%%%%%%%%%%%%%%%%%%%%%%%%%%%%%%%%%%%%%%%%%%%%%%%%%%%%%%%
%%%%%%%%%%%%%%%%%%%%%%%%%%%%%%%%%%%%%%%%%%%%%%%%%%%%%%%%%%%%%%%%%%%%%%%%%%%%%%%%%%%%%%%%%%%%%%%%%%%%%%%%
%%%%%%%%%%%%%%%%%%%%%%%%%%%%%%%%%%%%%%%%%%%%%%%%%%%%%%%%%%%%%%%%%%%%%%%%%%%%%%%%%%%%%%%%%%%%%%%%%%%%%%%%
%%%%%%%%%%%%%%%%%%%%%%%%%%%%%%%%%%%%%%%%%%%%%%%%%%%%%%%%%%%%%%%%%%%%%%%%%%%%%%%%%%%%%%%%%%%%%%%%%%%%%%%%

\section{The data}
\label{sec:data}
As we mentioned before, the data we used to compare all the classifiers were the same. The main characteristics of the data are described in this section.

The data were generated using the  simulation tool Pythia \cite{Sjostrand:2014zea}. Pythia is a toolkit for the generation of high-energy physics events, simulating, for instance, the proton collisions that occur at the LHC. Given that we use LHCb as the benchmark for our studies, we focused on several $B$ meson decay modes that were studied in the experiment. These correspond to different generic topologies and final-state particles. Note that we based our studies on the topology of the final states, and ignored any PID variable. LHCb has excellent libraries for PID, many of which rely on ML \cite{Kazeev:2744601}. We generated the following~decays:

\begin{itemize}

    \item $\bmu$
    \item $\bpipi$
    \item $\btripi$
    \item $\bcpi$
    
\end{itemize}

%XCV the verb needs to stay as it looks here or the environment is broken and excedes the line width, at least in my latex compiler
The procedure to generate the signal samples is as follows. We enabled the \verb|HardQCD:| \verb|gg2bbbard| and \verb|HardQCD:qqbar2bbbar| processes (which correspond to the production of a $b\bar{b}$ pair of quarks at the LHC) in Pythia at a collision energy of 14 TeV, and looked for $B$ mesons. We then redecayed these to our desired final state using the \verb|TGenPhaseSpace| tool from ROOT. Figure \ref{fig:simul} characterizes the usual signal topology, with a $B$ meson that flights a few mm and then decays to charged particles (such as pions or muons) whose trajectory and momentum can be fully reconstructed. We then applied the set of selection requirements included in Table~\ref{tab:cuts}. These are based on the quantities explained on Table~\ref{tab:features} and Figure~\ref{fig:simul}. On top of these quantities, we also applied selection requirements based on the pseudo-rapidity ($\eta$) of the final state particles. This relates to the angular acceptance of the detector, so that particles not falling in a specific $\eta$ range fall outside the detector and therefore cannot be reconstructed.

\begin{table}[H]
\centering
\small
\caption{Values for the cuts used in Pythia to generate the signal and background samples for the analysis. The meaning of these variables can be found in Table~\ref{tab:features} and Figure~\ref{fig:simul}, as well as in the main~text.\label{tab:cuts}}
%\centering
\setlength{\tabcolsep}{2.1mm}
%\resizebox{\myfigwidth \textwidth}{!}{
\begin{tabular}{lccccccc}
\toprule
      & \boldmath{$\eta$} &\boldmath{ $  p_{TB}$}  & \boldmath{$ p_{T\mu}, p_{T\pi} $}  &\boldmath{ $ IP_{B} $} &\boldmath{ $ IP_{\mu}, IP_{\pi} $} & \textbf{DOCA }& \textbf{DoF} \\ 
\midrule      
 $\mu$ & $2< \eta <5$ & $>$1000 MeV/c   & $>$500 MeV/c  & $<$0.1 mm & $>$0.5 mm & $<$ 0.1 mm & $>$3 mm   \\ 

  $\pi$ & $2< \eta <5$ & $>$1000 MeV/c &$>$500 MeV/c  & $<$0.5 mm& $>$0.5 mm & $<$0.1 mm & $>$3 mm     \\ 
\bottomrule
\end{tabular}%}
\end{table}
\vspace{-12pt}

\begin{table}[H]
\small
\centering
\caption{List of features used to build the classifiers. \label{tab:features} For a mathematical definition of quantities such as the DOCA or $IP$ see, e.g., Ref.~\cite{BuarqueFranzosi:2021kky}. More details are given in the text.}
%\resizebox{\myfigwidth \textwidth}{!}{ 
\setlength{\tabcolsep}{0.01mm}
  \begin{tabular}{cc}
  \midrule
     $ p_{TB}$ & Transverse momentum of the mother $B$ meson\\ 
     $ p_{Tdaug}$ & Transverse momentum of the daughter particles \\  
     $p_x,p_y,p_z$ & Momentum components  of the daughters\\  
     $IP_{B}$ & Closest distance between the $B$ mother trajectory and the proton--proton collision \\
      &vertex\\ 
     $IP_{daug}$& Closest distance between the daughter particle trajectory and the proton--proton \\
    & collision vertex\\  
     DOCA& Distance Of Closest Approach between daughter particles\\ 
     DoF& Distance of Flight between the production and decay points of the mother $B$ meson\\
     Isolation $\mu_{1}$& Minimum distance between $\mu_{1}$ and any particle produced by the $b\bar{b}$ pair,  excluding $\mu_{2}$\\ 
     Isolation $\mu_{2}$& same but for $\mu_{2}$, excluding $\mu_{1}$\\ 
     Daughters$_{\rm pos}$ & Position of daughter particles\\ 
     B$_{\rm pos}$ & Position of the $B$ particle\\ 
     \bottomrule
  \end{tabular}%}
  
\end{table}
\vspace{-12pt}
\begin{figure}[H]
    %\centering
    \ifthenelse{\boolean{isarxiv}}
    {\includegraphics[width=\textwidth]{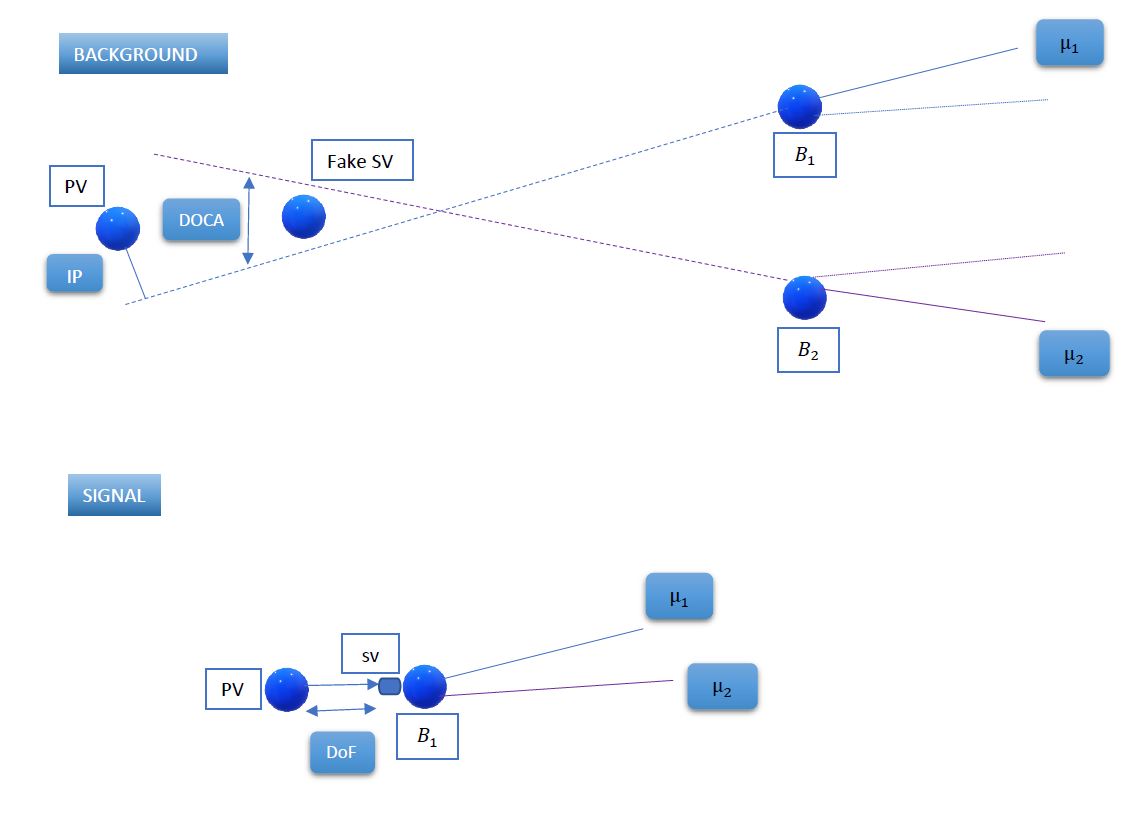}}
    {\includegraphics[width=0.7\textwidth]{images/all_example.PNG}}
    \caption{Characterization of the signal and background and definition of several variables to discriminate between them. For the $\mu\mu$ background, the usual candidates are formed by muons from different $B$ mesons that are incorrectly matched together. }
    \label{fig:simul}
\end{figure}

Figure~\ref{fig:simul} also explains the characterization of background events. We used the same type of Pythia processes to generate $b\bar{b}$ events, but there is no redecay at all this time, taking the event as it was originally fully simulated. We selected the charged pions and muons appearing in the event, grouped them, and applied the same cuts indicated  in Table~\ref{tab:cuts}. For some of the cuts, we needed to generate a fake $B$ meson decay vertex, which we used as the point in space minimizing the sum of distances to all the daughter particles.

In order to train the classifiers, we selected a list of variables (features) that was similar to those used to select the signal and background samples. These features are coherent between different types of classifiers to make the comparison fair. Note that some of the features chosen depend on the decay channel that was chosen. For instance, the $\bcpi$ decay has four particles in the final state, which means the $IP$ of two more particles is included compared to, e.g., $\bpipi$. As  explained below, different sub-selections of features were also tried to provide as complete a picture as possible. The full list of features that were used can be found in Table~\ref{tab:featuresapp}. An additional aspect we have accounted for is the detector resolution. Our simulation provides the ``true'' value of features, but real-life detectors have some inherent inaccuracies due to their resolutions, as discussed in the introduction, so what they measure differs from this; for instance, the resolution damages the discrimination power of some of the features we work with. While, in principle, this would be only be a problem when trying to find a realistic performance of the classifiers, our goal here is only to compare their performance. In any case, to use a dataset that is as close as possible to that used by LHCb analysts, we conducted a Gaussian smear of the variables that are most affected by resolution effects. These are the momenta of the particles, the DOCAs and $IPs$. To determine the associated resolutions, we based ourselves in References~\cite{CidVidal:2019qub,Hynds:2004399}.

Apart from the features introduced above, for the $\bmu$ case we added an additional variable, the isolation, which is known to provide excellent signal-background discrimination in analyses of this kind \cite{LHCb:2011arj}. The isolation exploits the fact that the usual $\bmu$ decays have nothing ``around'' other than the muons, while the background contains additional objects produced in $b-$hadron decays (see Figure~\ref{fig:simul}). To quantify this isolation, we calculated it as the minimum DOCA between our selected muons and every other charged particle in the event. These other particles were selected by applying the cuts in Table~\ref{tab:cutsiso}. Defined in this way, the isolation peaks at lower values for background and has larger peaks for the signal. %please check intended meaning has been retained
\begin{table}[H]
%\small
\centering
\caption{List of cuts applied to charged particles entering the computation of the isolation. See text for details.\label{tab:cutsiso}}
%\centering
\setlength{\tabcolsep}{14.65mm}
\begin{tabular}{ccc}
\toprule
\boldmath{$ \eta $} &\boldmath{$p_{T}$ }&\boldmath{ $IP$} \\ 
\midrule
$2< \eta <5$ &$>$250 MeV/c & $>$0.1 mm   \\ 
\bottomrule
\end{tabular}

\end{table}

The simulated data consist of around 3000 samples of both signal and background events, generated as explained above. A normal LHC analysis may face difficulties in finding enough signal or background events to train a classifier, for instance, due to the very small selection efficiencies that require the simulation of an unreasonable amount of events.  Additionally, analysts often face the choice between computing ``low''- or ``high''-level features. The former are those that were directly measured by the detector (for instance, the position or momentum or charged tracks), while the latter are combinations of those that are known to behave differently for the signal and backgrounds (such as the DOCA or $IP$, introduced above). While the usual classifiers typically rely on high-level features, a smart enough algorithm should  be as effective only by using the low-level ones. To make all of this more concrete, we create six different training combinations between features and the number of samples. These are all %Is the bold necessary? XCV removed
 features--high stats, all features--low stats, low-level features--high stats, low-level features--low stats, high-level features--high stats, high-level features--low stats. The meaning of these categories is as follows:

\begin{itemize}
    \item High stats: In this category, we use all the 3000 samples for both signal and background.
    \item Low stats: In this category, we use 30\% of the 3000 samples available for both signal and background.
    \item All features: In this category, we use all the features we have available for the data.
    \item High-level features: In this category, we only use high-level features.
    \item Low-level features: In this category, we only use low-level features.
\end{itemize}
With these categories, we aim to provide guidelines to analysts, helping them with different future analyses to choose the best tool to treat their data. 
All of the features used for each of the options mentioned above are shown in Appendix~\ref{app:highlow}. 

One important last aspect to discuss, before moving on to the final analyses, is how the classifiers that we build must be uncorrelated with the invariant mass. In particle physics, one typically builds the invariant mass (the name ``invariant'' arises from special relativity, since this mass is the same for any reference frame) out of the properties of the final state particles. For instance, for the $\bmu$ decay, if we know the four-momenta of the muons, we can determine the invariant mass of the $B$ mother. Since the value of the $B$ meson mass is known a priori, the invariant mass is an excellent way to separate signal from background events, since, for signal, the distribution of this variable peaks at that known value. Another advantage of the invariant mass is that its distribution can be parameterized in an analytical way, which allows ``counting'' the amount of signal and background events by performing statistical fits to data. Since the invariant mass is a common and widely used feature in analyses in particle physics, as a rule of thumb, it is very important that any classifier we build is not correlated with the invariant mass, since the final count of signal events is performed based on this variable. This is so much the case that efforts have been made to develop classifiers that are explicitly trained so that no correlation is produced with the invariant mass (or other designed external variable) \cite{Stevens:2013dya,Rogozhnikov:2014zea}. This is tricky, since a classifier might be able to learn the discrimination achieved by the invariant mass, provide an excellent AUC score and still not be useful for an actual analysis in particle physics. Accordingly, we followed a simple approach and ensured all our classifiers were not correlated with the invariant mass, which guarantees a fair comparison.

%%%%%%%%%%%%%%%%%%%%%%%%%%%%%%%%%%%%%%%%%%%%%%%%%%%%%%%%%%%%%%%%%%%%%%%%%%%%%%%%%%%%%%%%%%%%%%%%%%%%%%%%
%%%%%%%%%%%%%%%%%%%%%%%%%%%%%%%%%%%%%%%%%%%%%%%%%%%%%%%%%%%%%%%%%%%%%%%%%%%%%%%%%%%%%%%%%%%%%%%%%%%%%%%%
%%%%%%%%%%%%%%%%%%%%%%%%%%%%%%%%%%%%%%%%%%%%%%%%%%%%%%%%%%%%%%%%%%%%%%%%%%%%%%%%%%%%%%%%%%%%%%%%%%%%%%%%
%%%%%%%%%%%%%%%%%%%%%%%%%%%%%%%%%%%%%%%%%%%%%%%%%%%%%%%%%%%%%%%%%%%%%%%%%%%%%%%%%%%%%%%%%%%%%%%%%%%%%%%%
%%%%%%%%%%%%%%%%%%%%%%%%%%%%%%%%%%%%%%%%%%%%%%%%%%%%%%%%%%%%%%%%%%%%%%%%%%%%%%%%%%%%%%%%%%%%%%%%%%%%%%%%

\section{Scikit learn vs. TMVA }
\label{sec:tmva_sklearn}

As seen in Section~\ref{sec:ML_intro}, TMVA is the dominant reference tool for signal-background discrimination at LHC experiments. Therefore, in  this section we check, using the generic data presented above, whether TMVA provides the optimal discrimination when compared to Sklearn, assuming that one is using the same datasets and the same algorithm.
The exercises below are intended to show the same analysis being carried out with both TMVA and Sklearn libraries, with the goal of helping scientists to switch between the two according to their needs.

The main difference between Sklearn and TMVA is the way the data are processed. In Sklearn, we have to transform the ROOT \verb|TTrees| to a numpy array or a pandas dataframe in order to work with them. At present, libraries such as uproot \cite{jim_pivarski_2019_3504190} or rootnumpy~\cite{noel_dawe_2015_18806} allow us to make this change in the data format without extra effort. This option allows for theuse of other libraries, instead of TMVA, even for the usual datasets in LHC experiments, which tend to be ROOT \verb|TTrees|. 

Both TMVA and Sklearn offer a simple way of handling algorithms, allowing for easy changes to their hyperparameters and providing useful information about the training time or even the ranking of features. 
To analyse and explain how the use of different libraries could be advantageous to the users, in Figure~\ref{fig:tmva} and Table~\ref{tab:tmva}, we compare both libraries using BDT with AdaBoost for the $\bmu$ decay. In this case, we focus in the high-level features and high stats option for training. The comparison was made with the same hyperparameters and the same data for both libraries. This means that all the tunable parameters had the same value and the samples used for training and testing were the same for both cases. These hyperparameters were chosen as the best after performing a grid search, i.e., running through a range of the best score was reached.
The results of the classifiers can be seen in Figure~\ref{fig:tmva}, where Sklearn is shown to have better results. This is also shown in Table~\ref{tab:tmva}, where the AUC scores for $\bmu$ and the rest of the decays can be found.

\begin{figure}
%\centering
    \ifthenelse{\boolean{isarxiv}}
    {\includegraphics[width=\textwidth]{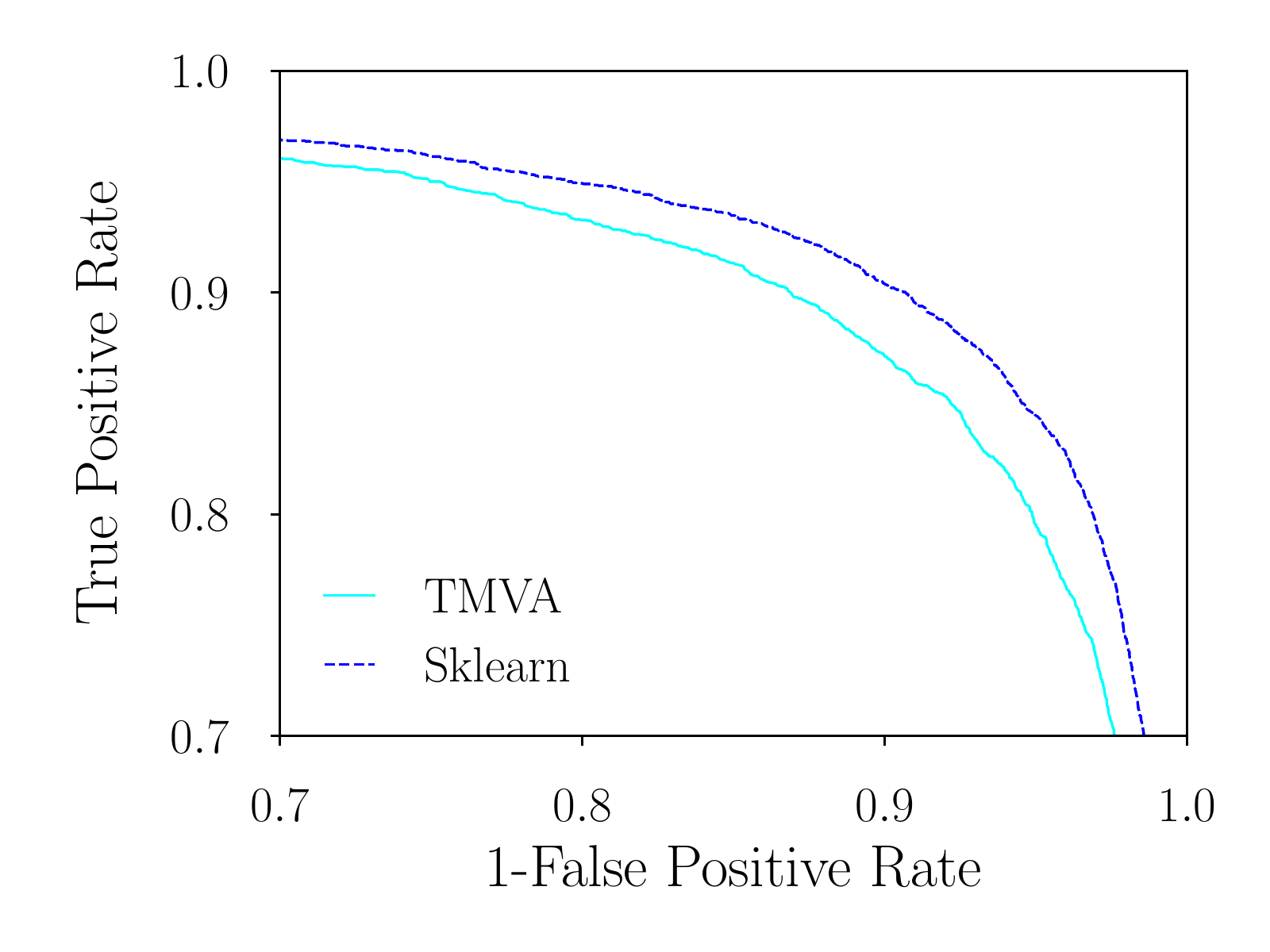}}
    {\includegraphics[width=0.7\textwidth]{images/tmva_zomm.pdf}}
     \caption{ROC Curve for the $\bmu$ decay, corresponding to the high-level features and high stats option for training.}
\label{fig:tmva}     
\end{figure}

\begin{table}[H]
\small
\centering
\caption{TMVA vs. Sklearn results: AUC scores for all four decay channels, corresponding to the high-level features and high stats option for training. The z value corresponds to a score \cite{delong,6851192} comparing the AUCs of pairs of classifiers under the hypothesis that they are the same. The corresponding \emph{p}-values are all $\sim 0$. \label{tab:tmva}}
%pasar a x 10^6 en latex
%\centering
\setlength{\tabcolsep}{9.25mm}
\begin{tabular}{llll}
\toprule
 & \textbf{TMVA AUC}  & \textbf{Sklearn AUC}  & \textbf{z Value}\\ 
 \midrule
$\bmu $ & 0.954   & 0.963   & 88.884\\ 
$\bpipi $ &  0.952   & 0.961  & 90.047\\ 
$\btripi $ &  0.957 & 0.960 &  88.224\\
$\bcpi $&  0.957  & 0.961  & 88.508\\ 
\bottomrule
  
\end{tabular}

\end{table}
In Table~\ref{tab:tmva}, we show all the results for the analyses that we defined in the previous section. For all these cases, Sklearn performs better than TMVA. This difference is small and not large enough to make a stronger statement, but it is still noticeable. To obtain a feeling of the statistical significance of the difference, we perform the DeLong test \cite{delong,6851192}, which allows for the AUCs of pairs of classifiers to be compared. This test provides a z score that helps to quantify the probability that the two classifiers perform differently when looking at the same test data. The larger the z score, the more significant the separation between both AUCs. When comparing Sklearn and TMVA, we find that, for all decay channels, the AUC scores are statistically significantly different. Note that this statistical test is limited to pairs of classifiers, so we do not apply it further in the paper, given the large amount of comparisons being performed in Section~\ref{sec:comparison}.

One of the reasons that we think could explain the differences we observe between libraries relates to how these are designed and have evolved.
 Sklearn was released in 2007, and TMVA in 2009.  Through the years, both libraries have changed and improved their performance until becoming those available at present. Still, as is shown in the tables and figures in this section, Sklearn always performs slightly better than TMVA. 
This means that even though both libraries a priori use the same algorithm, there are differences in the way these two frameworks operate. Different versions of these algorithms provide slightly different results, so one would be tempted to claim that Sklearn is able to provide a better optimization of the BDTs when compared to TMVA. This might be related to the application of the loss function or computational reasons.

%%%%%%%%%%%%%%%%%%%%%%%%%%%%%%%%%%%%%%%%%%%%%%%%%%%%%%%%%%%%%%%%%%%%%%%%%%%%%%%%%%%%%%%%%%%%%%%%%%%%%%%%
%%%%%%%%%%%%%%%%%%%%%%%%%%%%%%%%%%%%%%%%%%%%%%%%%%%%%%%%%%%%%%%%%%%%%%%%%%%%%%%%%%%%%%%%%%%%%%%%%%%%%%%%
%%%%%%%%%%%%%%%%%%%%%%%%%%%%%%%%%%%%%%%%%%%%%%%%%%%%%%%%%%%%%%%%%%%%%%%%%%%%%%%%%%%%%%%%%%%%%%%%%%%%%%%%
%%%%%%%%%%%%%%%%%%%%%%%%%%%%%%%%%%%%%%%%%%%%%%%%%%%%%%%%%%%%%%%%%%%%%%%%%%%%%%%%%%%%%%%%%%%%%%%%%%%%%%%%
%%%%%%%%%%%%%%%%%%%%%%%%%%%%%%%%%%%%%%%%%%%%%%%%%%%%%%%%%%%%%%%%%%%%%%%%%%%%%%%%%%%%%%%%%%%%%%%%%%%%%%%%

\section{Comparison of popular ML techniques beyond BDTs with AdaBoost }
\label{sec:comparison}

In this section, we show the results achieved using a wider range of algorithms and libraries for each of the decays explained in Section~\ref{sec:data}. The main libraries we use are Sklearn, PyTorch and Keras.  The main algorithms we use are BDT with AdaBoost and NNs, which were all introduced in Section~\ref{sec:ML_intro}. We limit ourselves to fully connected NNs with several~layers.

Although we began this exercise using both types of algorithms for each library, once we obtained the results, we realized that the results for BDT with AdaBoost, when compared to NNs, were worse for every analysis for the PyTorch and Keras libraries and every classification of the data. This is the reason that we show only the results for the NNs for these two libraries. Furthermore, since, in Section~\ref{sec:tmva_sklearn}, we found TMVA to have systematically lower AUC scores when compared to Sklearn, we did not include it in this second part. Finally, the results for Sklearn are shown using BDT with AdaBoost,, since these were almost the same as this library when compared to NNs. Therefore, we analysed a total of 72 different classifiers (4 decay channels, 6 training options and 3 libraries). Overall, the use NNs was shown to generally be as the most competitive option for the analyses in this paper. Note that, in each case, we performed a grid search to choose the hyperparameters with the best AUC scores. These were used later for the actual comparison. The values of the hyperparameters are given  in Appendix~\ref{sec:hyper}.

 We now show the results for all of the options we developed to select which is the best for each problem.
 Figures \ref{fig:bmu} and \ref{fig:bpi} show the results of all the previously mentioned analyses. For each column, we analysed a different decay and, for each row, we have a different training option for the analysis. 
 For example, in row 0 column 0, we show the ROC Curves for the $\bmu$ decay and the low-level features and low stats option, referred to as ``Low-Low''. Similar abbreviations appear for the other categories.

 \begin{figure}[H]
   % \centering
     \ifthenelse{\boolean{isarxiv}}
    {\includegraphics[width = 0.9\textwidth]{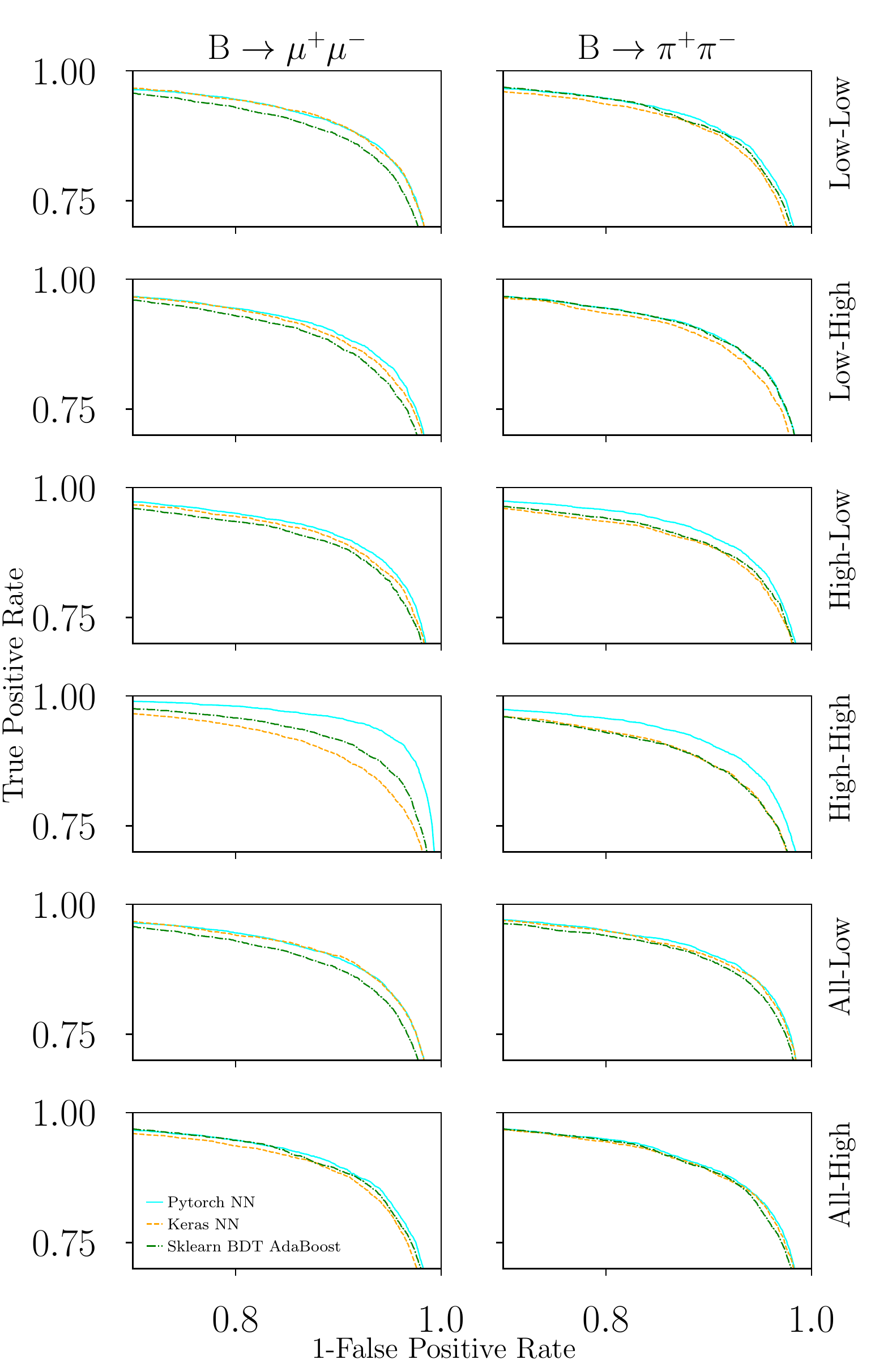}}
    {\includegraphics[width = 0.7\textwidth]{images/opcion1.pdf}}
        \caption{ROC Curve for the $\bmu$ and $\bpipi$ decays and the four different options for training. The options are those listed in Section~\ref{sec:data}.
         \label{fig:bmu}}
\end{figure}

 \begin{figure}[H]
 %   \centering
     \ifthenelse{\boolean{isarxiv}}
    {\includegraphics[width = 0.9\textwidth]{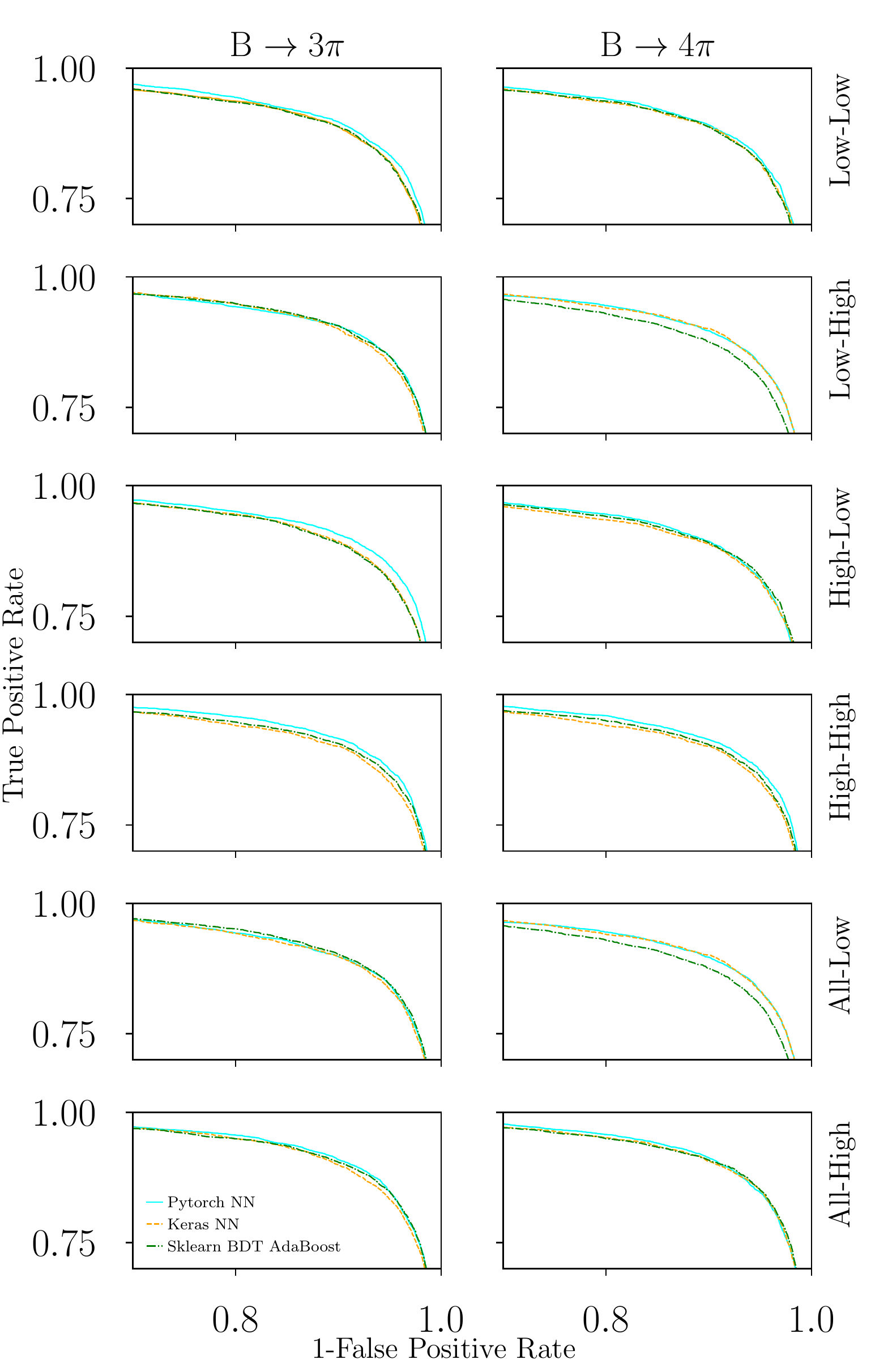}}
    {\includegraphics[width = 0.7\textwidth]{images/opcion2.pdf}}
         \caption{ \label{fig:bpi}ROC Curve for the $\btripi$ and $\bcpi$ decays and the four different options for training. The options are those listed in Section~\ref{sec:data}.}
\end{figure}

 %\newpage

%%%%%%%%%%%%%%%%%%%%%%%%%%%%%%%%%%%%%%%%%%%%%%%%%%%%%%%%%%%%%%%%%%%%%%%%%%%%%%%%%%%%%%%%%%%%%%%%%%%%%%%%
%%%%%%%%%%%%%%%%%%%%%%%%%%%%%%%%%%%%%%%%%%%%%%%%%%%%%%%%%%%%%%%%%%%%%%%%%%%%%%%%%%%%%%%%%%%%%%%%%%%%%%%%
%%%%%%%%%%%%%%%%%%%%%%%%%%%%%%%%%%%%%%%%%%%%%%%%%%%%%%%%%%%%%%%%%%%%%%%%%%%%%%%%%%%%%%%%%%%%%%%%%%%%%%%%
%%%%%%%%%%%%%%%%%%%%%%%%%%%%%%%%%%%%%%%%%%%%%%%%%%%%%%%%%%%%%%%%%%%%%%%%%%%%%%%%%%%%%%%%%%%%%%%%%%%%%%%%
%%%%%%%%%%%%%%%%%%%%%%%%%%%%%%%%%%%%%%%%%%%%%%%%%%%%%%%%%%%%%%%%%%%%%%%%%%%%%%%%%%%%%%%%%%%%%%%%%%%%%%%%

\section{Results and Discussion}
\label{sec:results}

Once we have calculated the ROC curves and obtained the AUC scores for each of the algorithms, we proceed to compare them all for each channel. In the following tables, the results are shown for each tool and each training option. Tables \ref{tab:bmumu}--\ref{tab:b4pi} follow the results for each analysis. For reference, some of the training times, which are comparable in all cases, can be found in Appendix~\ref{sec:time}. As shown in these tables, the AUC scores are similar for all of the options and most of the decays behave in a similar way, even when these scores are also subject to statistical fluctuations. The best results tend to be those obtained using PyTorch as the library, high-level features and all the available stats. This conclusion can be seen in Table~\ref{tab:final} and \ref{tab:best}. Here, we can appreciate how PyTorch is the dominant library for most of the options, as it is the $\bcpi $ decay, with all features and low stats and the $\bcpi$ decay low-level features and high stats serving as the two exceptions. %please check intended meaning has been retained
 In these cases, Keras worked better. However, the result for high-level features and high stats is still the best option for all the decays, as shown in Table~\ref{tab:best}. We note again that most of the results are consistent across different decay channels, which correspond to different topologies, different types of background and even slightly different features (e.g., the use of isolation for the $\bmu$ decay). 

When using the same decay and the same classification features, we can compare how differently the low-stat and high-stat options behave. In some cases, as in $\btripi$, we see how the result improves for all options when using high stats. This is the type of behaviour we would generally expect  to achieve when using ML. However, this is not something that happens in every case. For example, in the $\bpipi$ decay, we see that when using all features the AUC score does not improve, with more statistics, for Keras. The same occurs with Sklearn and high-level features and PyTorch and all features in the same channel. We observe that PyTorch and Sklearn tend to improve, with high stats, while Keras is more robust against a change. We also see a tendency to have a higher dependence on the statistics for the $\btripi$  and $\bcpi$ decays, which present more final state particles and, therefore, have more features to account for.

To check how the selection of features affects the performance of these algorithms, we note there are no large differences between different decay channels and between the low- and high-stats categories. We then count how often the best AUC score is obtained for the different libraries depending on the feature selection. One would a priori expect that the addition of information from secondary features at the ``all features'' option would benefit the discrimination, especially for NNs. This must be balanced by the fact that adding too many or redundant variables can damage the overall performance in some cases. When looking at the global picture, we see that PyTorch performs better with the high-level features, while for Keras, the results tend to be better when choosing all features. The situation is slightly more balanced for Sklearn, although all features tend to provide a worse performance. Additionally, the use of low-level features does not provide the best performance in every library, but this still does not dramatically affect the power of discrimination, which indicates that the algorithms we use are robust. %please check intended meaning has been retained
For reference, all the features used in this study can be found in Appendix~\ref{app:highlow}. Note that, as an additional element to judge the best choice, fewer features and smaller datasets can help to train the models faster and is  more manageable in terms of, e.g., grid searches. 

The convenience of using BDTs or NNs has long been a subject of discussion in the particle physics literature. For instance, Reference~\cite{ROE2005577} discusses how BDTs can improve the performance of NNs for PID in neutrino detectors. Note that these results do not include some of the latest progress concerning the training of NNs, which provided essential improvements in the last decade \cite{wang}. It is is remarkable how these improvements recover NNs as the best solution in the same dataset \cite{Stanev:2021mkr}. Other examples, which are more related to the type of analysis we present here, look into different algorithms in datasets that are often used by the ATLAS and CMS experiments, finding consistently better or similar results when comparing NNs to BDTs with different types of boosting, such as XGBoost \cite{Alvestad:2021sje,Tannenwald:2020mhq}. One of the first attempts to apply QML in a data sample comparable to ours, composed of $B$ meson decays \cite{Heredge:2021vww}, is also notable. The AUC achieved in this study manages to improve on those of other classical methods, although no NNs are included. Reference~\cite{Terashi:2020wfi}  compares QML algorithms to NNs in a different HEP dataset, achieving better or similar AUC scores, although this depends on the size of the training sample. Another interesting example concerns the use of these algorithms for the simulation of HEP data \cite{Bendavid:2017zhk}, where, again, NNs beat BDTs in terms of regression. Regarding the libraries used, the advantages of using PyTorch to deal with NNs are becoming well known, and several toolkits based on it have recently been developed for different HEP applications. Examples include Lumin \cite{Strong:2020mge,lumin} and Ariadne \cite{Goncharov:2021wvd}. The first is a wrapper that includes some of the newest techniques that facilitate the training of NNs, while the second, as mentioned in the introduction, is used for tracking. As a final remark, the benefits NNs can bring over other methods, such as BDTs, not only concern physics, but also appear in fields such as medicine \cite{8037515,abdar}, finance \cite{alexei}, marketing \cite{chaochao}, biology~\cite{partin} or engineering \cite{water}. As a general trend, NNs tend to perform better in these references, although the present developments in, e.g., BDTs, make these very competitive. We note the interest in looking at other sources, beyond particle physics, to find inspiration and discover better models for signal classification.

\begin{table}[H]
%\small
\centering
\caption{$\bmu$ channel: AUC scores for the different classifiers with all the libraries. \label{tab:bmumu}}
%\centering
\setlength{\tabcolsep}{7.25mm}
\begin{tabular}{lllll}
\toprule
  \boldmath{$\bmu$}   & \textbf{PyTorch}  & \textbf{Keras} & \textbf{Sklearn}  \\ 
  \midrule
 low-level features--low stats & 0.961  &  0.959   &   0.952    \\
 low-level features--high stats   &  0.959   &  0.958   &  0.953\\
 high-level features--low stats  &  0.964    &  0.959    &  0.956\\ 
 high-level features--high stats &  0.981    &  0.958    &  0.967\\
 all features--low stats  &  0.961    &  0.960    &  0.953   \\ 
 all features--high stats &  0.961    &  0.954   &  0.960 \\
 \bottomrule
\end{tabular}
\end{table}
\vspace{-12pt}

%%%%
\begin{table}[H]
%\small
\centering
\caption{$\bpipi$ channel: AUC scores for the different classifiers with all the libraries. \label{tab:bpipi}}
%\centering
%\resizebox{\myfigwidth \textwidth}{!}{
\setlength{\tabcolsep}{7.25mm}
\begin{tabular}{lllll}
\toprule
   \boldmath{$\bpipi$}   & \textbf{PyTorch}  & \textbf{Keras} & \textbf{Sklearn} \\ 
\midrule
 low-level features--low stats &  0.960    &    0.954  &   0.959    \\ 
  low-level features--high stats   & 0.960   &  0.957    &  0.960\\ 
 high-level features--low stats  & 0.965   & 0.956  &  0.958 \\
  high-level features--high stats &0.967 &0.954 & 0.954   \\
 all features--low stats  &  0.963    &  0.962    &  0.959   \\ 
 all features--high stats &  0.961   &  0.960    &  0.960 \\
 \bottomrule
\end{tabular}%}
\end{table}

\vspace{-12pt}

\begin{table}[H]
\centering
\caption{$\btripi$ channel: AUC scores for the different classifiers with all the libraries. \label{tab:b3pi}}
%\resizebox{\myfigwidth \textwidth}{!}{
\setlength{\tabcolsep}{7.25mm}
\begin{tabular}{lllll}
\toprule
 \boldmath{$\btripi$}  & \textbf{PyTorch} & \textbf{Keras} & \textbf{Sklearn}  \\ 
   \midrule
 low-level features--low stats &  0.961    &  0.954    &  0.956\\ 
 low-level features--high stats   & 0.963 &0.961 & 0.963  \\ 
 high-level features--low stats  &0.965  & 0.960 & 0.953  \\ 
 high-level features--high stats &  0.967 & 0.960  &  0.963   \\
  all features--low stats  &  0.961    &  0.960    &  0.963    \\ 
 all features--high stats &  0.965   &  0.962    &  0.963 \\
\bottomrule
\end{tabular}%}
\end{table}
\vspace{-12pt}

\begin{table}[H]
%\small
\centering
\caption{$\bcpi$ channel: AUC scores for the different classifiers with all the libraries. \label{tab:b4pi}}
%\centering
%\resizebox{\myfigwidth \textwidth}{!}{
\setlength{\tabcolsep}{7.25mm}
\begin{tabular}{lllll}
\toprule
    \boldmath{$\bcpi$}  & \textbf{PyTorch} & \textbf{Keras} & \textbf{Sklearn}  \\ 
   \midrule
 low-level features--low stats &  0.958    &  0.956 &  0.955 \\ 
 low-level features--high stats   & 0.961 & 0.961 & 0.953   \\ 
 high-level features--low stats  & 0.960 & 0.958 & 0.958  \\ 
 high-level features--high stats & 0.967  & 0.961 &  0.964  \\ 
  all features--low stats  &  0.960   &  0.961   &  0.954    \\ 
 all features--high stats &  0.967    &  0.964   &  0.964 \\
\bottomrule
\end{tabular}%}
\end{table}
\vspace{-12pt}

\begin{table}[H]
%\small
\centering
\caption{Best library for each analysis and each training option for the data. \label{tab:final}}
%\centering
%\resizebox{\mytabwidth \textwidth}{!}{
\setlength{\tabcolsep}{2.65mm}
\begin{tabular}{lllll}
\toprule
     &  \boldmath{$\bmu $} &  \boldmath{$\bpipi$ }&  \boldmath{ $\btripi $}  & \boldmath{ $\bcpi$}  \\
\midrule
 low-level features--low stats &   PyTorch   &  PyTorch    & PyTorch     & PyTorch  \\ 
 low-level features--high stats   &  PyTorch    &  PyTorch    &  PyTorch      & Keras  \\ 
 high-level features--low stats  &  PyTorch    &  PyTorch    &  PyTorch    & PyTorch  \\ 
 high-level features--high stats &  PyTorch     &  PyTorch     &  PyTorch    & PyTorch\\ 
 all features--low stats  &  PyTorch   &  PyTorch    &  PyTorch  & Keras \\ 
 all features--high stats &  PyTorch    &  PyTorch    &  PyTorch &  PyTorch\\
 \bottomrule
\end{tabular}%}
\end{table}
\vspace{-12pt}

\begin{table}[H]
%\small
\centering
\caption{Best training option for the data and library for each analysis. \label{tab:best}}
%\centering
%\resizebox{\myfigwidth \textwidth}{!}{
\setlength{\tabcolsep}{9.65mm}
\begin{tabular}{lll}
\toprule
 $\bmu $ & high-level features--high stats & PyTorch  \\ 
 $\bpipi$ & high-level features--high stats &  PyTorch    \\
$\btripi$ & high-level features--high stats&  PyTorch     \\ 
$\bcpi$& high-level features--high stats &  PyTorch    \\ 
\bottomrule
\end{tabular}%}
\end{table}

%%%%%%%%%%%%%%%%%%%%%%%%%%%%%%%%%%%%%%%%%%%%%%%%%%%%%%%%%%%%%%%%%%%%%%%%%%%%%%%%%%%%%%%%%%%%%%%%%%%%%%%%
%%%%%%%%%%%%%%%%%%%%%%%%%%%%%%%%%%%%%%%%%%%%%%%%%%%%%%%%%%%%%%%%%%%%%%%%%%%%%%%%%%%%%%%%%%%%%%%%%%%%%%%%
%%%%%%%%%%%%%%%%%%%%%%%%%%%%%%%%%%%%%%%%%%%%%%%%%%%%%%%%%%%%%%%%%%%%%%%%%%%%%%%%%%%%%%%%%%%%%%%%%%%%%%%%
%%%%%%%%%%%%%%%%%%%%%%%%%%%%%%%%%%%%%%%%%%%%%%%%%%%%%%%%%%%%%%%%%%%%%%%%%%%%%%%%%%%%%%%%%%%%%%%%%%%%%%%%
%%%%%%%%%%%%%%%%%%%%%%%%%%%%%%%%%%%%%%%%%%%%%%%%%%%%%%%%%%%%%%%%%%%%%%%%%%%%%%%%%%%%%%%%%%%%%%%%%%%%%%%%

\section{Conclusions}
\label{sec:conclusions}

In this paper, we review the most frequent ML libraries used in HEP, checking their popularity and comparing their performance in terms of signal-background discrimination in an independent simulated dataset. These datasets are generated and selected using the LHCb detector at CERN as a benchmark. 

In the first part of the paper, we observe that, even though TMVA is one of the most popular libraries, and by  far the more frequently used for this type of analyses, there are other options that can work as well or even better. In fact, Sklearn performs better than TMVA in all the decays analysed in this paper when using BDTs boosted with AdaBoost. This can be a way to prompt scientists in HEP collaborations to try new alternatives and generate new content using the most modern libraries that are available according to Python and other languages. Thanks to the wide range of new libraries, the conversion between the ROOT files and Python-friendly data structures is growing easier over time, and this can result in the popularization of these modern ML libraries for particle physicists.

In the second part of the paper, we compared the obtained results with some of the most popular ML libraries in data science, namely, Keras, PyTorch and Sklearn, and showed how NNs can improve the results obtained by BDTs. Even if the results are similar between the three of them, PyTorch tends to provide the best scores. In any case, the final choice might depend on other aspects, such as the amount of statistics available, since, for instance, Keras appears to be more robust for lower training statistics. Regarding the selection of features, this depends on the library, but, as a rule of thumb, we recommend not enlarging the list unnecessarily, focusing on high-level libraries that are known to provide excellent signal-background discrimination. %please check intended meaning has been retained
To our knowledge, this is the first time that a detailed study of the dependence of algorithms and libraries on the training statistics and number of features has been performed with an HEP dataset. 

We should highlight that, even though we are able to see some differences among the libraries, algorithms and decays, all AUC scores are in the range of 0.95--0.96, and subject to statistical fluctuations. This means that the results we have obtained are indicative, but must still be made with caution. Therefore, even if our findings can be used as a guideline, we still advise analysts to check for the best classifier for their specific case, depending on aspects such as available statistics or the available CPU for training.  

As a final note, we emphasize  the need for particle physicists to enrich their perspective by looking at what is being carried out in other fields in terms of classification by means of ML. This not only concerns other areas of physics \cite{Carleo:2019ptp}, but also fields that look far away, such as medicine \cite{Rajkomar2019-vl}, chemistry \cite{keith}, industrial applications \cite{QIAN2021100642,NASIRI20211137} or the interface with users \cite{Dudley2018ARO}.

%%%%%%%%%%%%%%%%%%%%%%%%%%%%%%%%%%%%%%%%%%%%%%%%%%%%%%%%%%%%%%%%%%%%%%%%%%%%%%%%%%%%%%%%%%%%%%%%%%%%%%%%
%%%%%%%%%%%%%%%%%%%%%%%%%%%%%%%%%%%%%%%%%%%%%%%%%%%%%%%%%%%%%%%%%%%%%%%%%%%%%%%%%%%%%%%%%%%%%%%%%%%%%%%%
%%%%%%%%%%%%%%%%%%%%%%%%%%%%%%%%%%%%%%%%%%%%%%%%%%%%%%%%%%%%%%%%%%%%%%%%%%%%%%%%%%%%%%%%%%%%%%%%%%%%%%%%
%%%%%%%%%%%%%%%%%%%%%%%%%%%%%%%%%%%%%%%%%%%%%%%%%%%%%%%%%%%%%%%%%%%%%%%%%%%%%%%%%%%%%%%%%%%%%%%%%%%%%%%%
%%%%%%%%%%%%%%%%%%%%%%%%%%%%%%%%%%%%%%%%%%%%%%%%%%%%%%%%%%%%%%%%%%%%%%%%%%%%%%%%%%%%%%%%%%%%%%%%%%%%%%%%

\section*{Acknowledgements}

We thank Alexandre Brea and Titus Mombacher for reading our manuscript and providing useful comments.

This work has received financial support from Xunta de Galicia (Centro singular de investigaci\'on de Galicia accreditation 2019-2022), by European Union ERDF, and by  the ``Mar\'ia  de Maeztu''  Units  of  Excellence program  MDM-2016-0692 and  the Spanish Research State Agency. In particular: the work of X.C.V. is supported by MINECO (Spain) through the Ram\'{o}n y Cajal program RYC-2016-20073 and by XuntaGAL under the ED431F 2018/01 project and the work of L.D.M is supported by the Spanish Research State Agency (Spain) through the ``Doctorados Industriales'' program DIN2018-010092B70548706.

\bibliographystyle{JHEP}
\bibliography{main}

\providecommand{\href}[2]{#2}\begingroup\raggedright\begin{thebibliography}{10}

\bibitem{Feickert:2021ajf}
M.~Feickert and B.~Nachman, \emph{{A Living Review of Machine Learning for
  Particle Physics}},  \href{https://arxiv.org/abs/2102.02770}{{\ttfamily
  2102.02770}}.

\bibitem{LHCb:2008vvz}
{\scshape $\rm LHCb$} collaboration, \emph{{The LHCb Detector at the LHC}},
  \href{https://doi.org/10.1088/1748-0221/3/08/S08005}{\emph{JINST} {\bfseries
  3} (2008) S08005}.

\bibitem{LHCb:2014set}
{\scshape $\rm LHCb$} collaboration, \emph{{LHCb Detector Performance}},
  \href{https://doi.org/10.1142/S0217751X15300227}{\emph{Int. J. Mod. Phys. A}
  {\bfseries 30} (2015) 1530022}
  [\href{https://arxiv.org/abs/1412.6352}{{\ttfamily 1412.6352}}].

\bibitem{cern_data}
``{CERN storage}.'' \url{https://home.cern/science/computing/storage}.

\bibitem{Qasim:2019otl}
S.R.~Qasim, J.~Kieseler, Y.~Iiyama and M.~Pierini, \emph{{Learning
  representations of irregular particle-detector geometry with
  distance-weighted graph networks}},
  \href{https://doi.org/10.1140/epjc/s10052-019-7113-9}{\emph{Eur. Phys. J. C}
  {\bfseries 79} (2019) 608}
  [\href{https://arxiv.org/abs/1902.07987}{{\ttfamily 1902.07987}}].

\bibitem{Cranmer:2021gdt}
K.~Cranmer, M.~Drnevich, S.~Macaluso and D.~Pappadopulo, \emph{{Reframing Jet
  Physics with New Computational Methods}},
  \href{https://doi.org/10.1051/epjconf/202125103059}{\emph{EPJ Web Conf.}
  {\bfseries 251} (2021) 03059}
  [\href{https://arxiv.org/abs/2105.10512}{{\ttfamily 2105.10512}}].

\bibitem{Goncharov:2021wvd}
P.~Goncharov, E.~Schavelev, A.~Nikolskaya and G.~Ososkov, \emph{{Ariadne:
  PyTorch Library for Particle Track Reconstruction Using Deep Learning}},
  \href{https://doi.org/10.1063/5.0063300}{\emph{AIP Conf. Proc.} {\bfseries
  2377} (2021) 040004} [\href{https://arxiv.org/abs/2109.08982}{{\ttfamily
  2109.08982}}].

\bibitem{Andrews:2021ejw}
M.~Andrews et~al., \emph{{End-to-End Jet Classification of Boosted Top Quarks
  with CMS Open Data}},
  \href{https://doi.org/10.1051/epjconf/202125104030}{\emph{EPJ Web Conf.}
  {\bfseries 251} (2021) 04030}
  [\href{https://arxiv.org/abs/2104.14659}{{\ttfamily 2104.14659}}].

\bibitem{Akchurin:2021ahx}
N.~Akchurin, C.~Cowden, J.~Damgov, A.~Hussain and S.~Kunori,
  \emph{{Perspectives on the Calibration of CNN Energy Reconstruction in Highly
  Granular Calorimeters}},  \href{https://arxiv.org/abs/2108.10963}{{\ttfamily
  2108.10963}}.

\bibitem{Cornell:2021gut}
A.S.~Cornell, W.~Doorsamy, B.~Fuks, G.~Harmsen and L.~Mason, \emph{{Boosted
  decision trees in the era of new physics: a smuon analysis case study}},
  \href{https://arxiv.org/abs/2109.11815}{{\ttfamily 2109.11815}}.

\bibitem{Baldi:2014kfa}
P.~Baldi, P.~Sadowski and D.~Whiteson, \emph{{Searching for Exotic Particles in
  High-Energy Physics with Deep Learning}},
  \href{https://doi.org/10.1038/ncomms5308}{\emph{Nature Commun.} {\bfseries 5}
  (2014) 4308} [\href{https://arxiv.org/abs/1402.4735}{{\ttfamily 1402.4735}}].

\bibitem{Dillon:2020quc}
B.M.~Dillon, D.A.~Faroughy, J.F.~Kamenik and M.~Szewc, \emph{{Learning the
  latent structure of collider events}},
  \href{https://doi.org/10.1007/JHEP10(2020)206}{\emph{JHEP} {\bfseries 10}
  (2020) 206} [\href{https://arxiv.org/abs/2005.12319}{{\ttfamily
  2005.12319}}].

\bibitem{Dahbi:2020zjw}
S.-e.~Dahbi, J.~Choma, B.~Mellado, G.~Mokgatitswane, X.~Ruan, B.~Lieberman
  et~al., \emph{{Machine learning approach for the search of resonances with
  topological features at the Large Hadron Collider}},
  \href{https://arxiv.org/abs/2011.09863}{{\ttfamily 2011.09863}}.

\bibitem{LHCb:2011arj}
{\scshape $\rm LHCb$} collaboration, \emph{{Search for the rare decays $B^0_s
  \to \mu^+\mu^-$ and $B^0 \to \mu^+\mu^-$}},
  \href{https://doi.org/10.1016/j.physletb.2011.04.031}{\emph{Phys. Lett. B}
  {\bfseries 699} (2011) 330}
  [\href{https://arxiv.org/abs/1103.2465}{{\ttfamily 1103.2465}}].

\bibitem{Williams:1323557}
M.~Williams, V.V.~Gligorov, C.~Thomas, H.~Dijkstra, J.~Nardulli and
  P.~Spradlin, \emph{{The HLT2 Topological Lines}},  Tech. Rep.
  \href{https://cds.cern.ch/record/1323557}{}, CERN, Geneva (Jan, 2011).

\bibitem{Likhomanenko:2015aba}
T.~Likhomanenko, P.~Ilten, E.~Khairullin, A.~Rogozhnikov, Ustyuzhanin et~al.,
  \emph{{LHCb Topological Trigger Reoptimization}},
  \href{https://doi.org/10.1088/1742-6596/664/8/082025}{\emph{J. Phys. Conf.
  Ser.} {\bfseries 664} (2015) 082025}
  [\href{https://arxiv.org/abs/1510.00572}{{\ttfamily 1510.00572}}].

\bibitem{higgs_dataset}
C.~Adam-Bourdarios, G.~Cowan, C.~Germain, I.~Guyon, B.~K\'{e}gl and
  D.~Rousseau, \emph{The higgs boson machine learning challenge},  in
  \emph{Proceedings of the 2014 International Conference on High-Energy Physics
  and Machine Learning - Volume 42}, HEPML'14, p.~19–55, JMLR.org, 2014.

\bibitem{Aarrestad:2021oeb}
T.~Aarrestad et~al., \emph{{The Dark Machines Anomaly Score Challenge:
  Benchmark Data and Model Independent Event Classification for the Large
  Hadron Collider}},  \href{https://arxiv.org/abs/2105.14027}{{\ttfamily
  2105.14027}}.

\bibitem{Kasieczka:2021xcg}
G.~Kasieczka et~al., \emph{{The LHC Olympics 2020: A Community Challenge for
  Anomaly Detection in High Energy Physics}},
  \href{https://arxiv.org/abs/2101.08320}{{\ttfamily 2101.08320}}.

\bibitem{TMVA:1019880}
A.~Hocker, P.~Speckmayer, J.~Stelzer, J.~Therhaag, E.~von Toerne, H.~Voss
  et~al., \emph{{TMVA - Toolkit for Multivariate Data Analysis with ROOT: Users
  guide. TMVA - Toolkit for Multivariate Data Analysis}},  Tech. Rep.
  \href{https://cds.cern.ch/record/1019880}{}, CERN, Geneva (Mar, 2007).

\bibitem{rene_brun_2019_3895860}
R.~Brun, F.~Rademakers, P.~Canal, A.~Naumann, O.~Couet, L.~Moneta et~al.,
  \emph{root-project/root: v6.18/02}, .

\bibitem{pedregosa2011scikit}
F.~Pedregosa, G.~Varoquaux, A.~Gramfort, V.~Michel, B.~Thirion et~al.,
  \emph{Scikit-learn: Machine learning in python}, {\emph{Journal of machine
  learning research} {\bfseries 12} (2011) 2825}.

\bibitem{Geron2017Handson}
A.~G\'{e}ron, \emph{Hands-on machine learning with {Scikit-Learn} and
  {TensorFlow} concepts, tools, and techniques to build intelligentsystems},
  O'Reilly Media, Apr., 2017,
  \href{http://www.amazon.com/exec/obidos/redirect?tag=citeulike07-20\&path=ASIN/1491962291}{http://www.amazon.com/exec/obidos/redirect?tag=citeulike07-20\&path=ASIN/1491962291}.

\bibitem{Hunter:2007}
J.D.~Hunter, \emph{Matplotlib: A 2d graphics environment},
  \href{https://doi.org/10.1109/MCSE.2007.55}{\emph{Computing in Science \&
  Engineering} {\bfseries 9} (2007) 90}.

\bibitem{mckinney2010data}
W.~McKinney et~al., \emph{Data structures for statistical computing in python},
   in \emph{Proceedings of the 9th Python in Science Conference}, vol.~445,
  pp.~51--56, Austin, TX, 2010.

\bibitem{2020SciPy-NMeth}
P.~Virtanen, R.~Gommers, T.E.~Oliphant, M.~Haberland, T.~Reddy, D.~Cournapeau
  et~al., \emph{{{SciPy} 1.0: Fundamental Algorithms for Scientific Computing
  in Python}}, \href{https://doi.org/10.1038/s41592-019-0686-2}{\emph{Nature
  Methods} {\bfseries 17} (2020) 261}.

\bibitem{harris2020array}
C.R.~Harris, K.J.~Millman, S.J.~van~der Walt, R.~Gommers, P.~Virtanen,
  D.~Cournapeau et~al., \emph{Array programming with {NumPy}},
  \href{https://doi.org/10.1038/s41586-020-2649-2}{\emph{Nature} {\bfseries
  585} (2020) 357}.

\bibitem{NEURIPS2019_9015}
A.~Paszke, S.~Gross, F.~Massa, A.~Lerer, J.~Bradbury, G.~Chanan et~al.,
  \emph{Pytorch: An imperative style, high-performance deep learning library},
  in \emph{Advances in Neural Information Processing Systems 32}, H.~Wallach,
  H.~Larochelle, A.~Beygelzimer, F.~d\textquotesingle Alch\'{e}-Buc, E.~Fox and
  R.~Garnett, eds., pp.~8024--8035, Curran Associates, Inc. (2019),
  \href{http://papers.neurips.cc/paper/9015-pytorch-an-imperative-style-high-performance-deep-learning-library.pdf}{http://papers.neurips.cc/paper/9015-pytorch-an-imperative-style-high-performance-deep-learning-library.pdf}.

\bibitem{chollet2015keras}
F.~Chollet et~al., \emph{Keras},  GitHub (2015),
  \href{https://github.com/fchollet/keras}{https://github.com/fchollet/keras}.

\bibitem{tensorflow2015-whitepaper}
M.~Abadi, A.~Agarwal, P.~Barham, E.~Brevdo, Z.~Chen, C.~Citro et~al.,
  \emph{{TensorFlow}: Large-scale machine learning on heterogeneous systems},
  2015.

\bibitem{2016arXiv160502688full}
{\scshape Theano Development Team} collaboration, \emph{{Theano: A {Python}
  framework for fast computation of mathematical expressions}}, {\emph{arXiv
  e-prints} {\bfseries abs/1605.02688} (2016) }.

\bibitem{10.1145/2939672.2945397}
F.~Seide and A.~Agarwal, \emph{Cntk: Microsoft's open-source deep-learning
  toolkit},  in \emph{Proceedings of the 22nd ACM SIGKDD International
  Conference on Knowledge Discovery and Data Mining}, KDD '16, (New York, NY,
  USA), p.~2135, Association for Computing Machinery, 2016,
  \href{https://doi.org/10.1145/2939672.2945397}{DOI}.

\bibitem{DENBY1988429}
B.~Denby, \emph{Neural networks and cellular automata in experimental high
  energy physics},
  \href{https://doi.org/https://doi.org/10.1016/0010-4655(88)90004-5}{\emph{Computer
  Physics Communications} {\bfseries 49} (1988) 429}.

\bibitem{xgboost}
T.~Chen and C.~Guestrin, \emph{Xgboost},
  \href{https://doi.org/10.1145/2939672.2939785}{\emph{Proceedings of the 22nd
  ACM SIGKDD International Conference on Knowledge Discovery and Data Mining}
  (2016) }.

\bibitem{icax}
J.~Duan, P.~Asteris, H.~Nguyen, X.-N.~Bui and H.~Moayedi, \emph{A novel
  artificial intelligence technique to predict compressive strength of recycled
  aggregate concrete using ica-xgboost model},
  \href{https://doi.org/10.1007/s00366-020-01003-0}{\emph{Engineering With
  Computers} {\bfseries 37} (2021) }.

\bibitem{woa}
Y.~Qiu, J.~Zhou, M.~Khandelwal, H.~Yang, P.~Yang and C.~Li, \emph{Performance
  evaluation of hybrid woa-xgboost, gwo-xgboost and bo-xgboost models to
  predict blast-induced ground vibration},
  \href{https://doi.org/10.1007/s00366-021-01393-9}{\emph{Engineering with
  Computers} (2021) 1}.

\bibitem{Hasani_Lechner_Amini_Rus_Grosu_2021}
R.~Hasani, M.~Lechner, A.~Amini, D.~Rus and R.~Grosu, \emph{Liquid
  time-constant networks}, {\emph{Proceedings of the AAAI Conference on
  Artificial Intelligence} {\bfseries 35} (2021) 7657}.

\bibitem{qml}
M.~Schuld, I.~Sinayskiy and F.~Petruccione, \emph{An introduction to quantum
  machine learning},
  \href{https://doi.org/10.1080/00107514.2014.964942}{\emph{Contemporary
  Physics} {\bfseries 56} (2014) 172–185}.

\bibitem{ATLAS}
{\scshape ATLAS} collaboration, \emph{{The ATLAS Experiment at the CERN Large
  Hadron Collider}},
  \href{https://doi.org/10.1088/1748-0221/3/08/S08003}{\emph{JINST} {\bfseries
  3} (2008) S08003}.

\bibitem{CMS}
{\scshape CMS} collaboration, \emph{{The CMS Experiment at the CERN LHC}},
  \href{https://doi.org/10.1088/1748-0221/3/08/S08004}{\emph{JINST} {\bfseries
  3} (2008) S08004}.

\bibitem{Sjostrand:2014zea}
T.~Sj\"ostrand, S.~Ask, J.R.~Christiansen, R.~Corke, Desai and others.,
  \emph{{An introduction to PYTHIA 8.2}},
  \href{https://doi.org/10.1016/j.cpc.2015.01.024}{\emph{Comput. Phys. Commun.}
  {\bfseries 191} (2015) 159}
  [\href{https://arxiv.org/abs/1410.3012}{{\ttfamily 1410.3012}}].

\bibitem{Kazeev:2744601}
N.~Kazeev, \emph{{Machine Learning for particle identification in the LHCb
  detector}}, .

\bibitem{BuarqueFranzosi:2021kky}
D.~Buarque~Franzosi, G.~Cacciapaglia, X.~Cid~Vidal, G.~Ferretti, T.~Flacke and
  C.~V\'azquez~Sierra, \emph{{Exploring new possibilities to discover a light
  pseudo-scalar at LHCb}},  \href{https://arxiv.org/abs/2106.12615}{{\ttfamily
  2106.12615}}.

\bibitem{CidVidal:2019qub}
X.~Cid~Vidal, P.~Ilten, J.~Plews, B.~Shuve and Y.~Soreq, \emph{{Discovering
  True Muonium at LHCb}},
  \href{https://doi.org/10.1103/PhysRevD.100.053003}{\emph{Phys. Rev. D}
  {\bfseries 100} (2019) 053003}
  [\href{https://arxiv.org/abs/1904.08458}{{\ttfamily 1904.08458}}].

\bibitem{Hynds:2004399}
D.P.M.~Hynds, \emph{{Resolution studies and performance evaluation of the LHCb
  VELO upgrade}}, .

\bibitem{Stevens:2013dya}
J.~Stevens and M.~Williams, \emph{{uBoost: A boosting method for producing
  uniform selection efficiencies from multivariate classifiers}},
  \href{https://doi.org/10.1088/1748-0221/8/12/P12013}{\emph{JINST} {\bfseries
  8} (2013) P12013} [\href{https://arxiv.org/abs/1305.7248}{{\ttfamily
  1305.7248}}].

\bibitem{Rogozhnikov:2014zea}
A.~Rogozhnikov, A.~Bukva, V.V.~Gligorov, A.~Ustyuzhanin and M.~Williams,
  \emph{{New approaches for boosting to uniformity}},
  \href{https://doi.org/10.1088/1748-0221/10/03/T03002}{\emph{JINST} {\bfseries
  10} (2015) T03002} [\href{https://arxiv.org/abs/1410.4140}{{\ttfamily
  1410.4140}}].

\bibitem{jim_pivarski_2019_3504190}
J.~Pivarski, P.~Das, D.~Smirnov, C.~Burr, M.~Feickert, N.~Biederbeck et~al.,
  \emph{scikit-hep/uproot: 3.10.7}, .

\bibitem{noel_dawe_2015_18806}
N.~Dawe, P.~Ongmongkolkul, C.~Deil, G.~Stark, P.~Waller, J.~Howard et~al.,
  \emph{root\_numpy: 4.2.0}, .

\bibitem{delong}
E.R.~DeLong, D.M.~DeLong and D.L.~Clarke-Pearson, \emph{Comparing the areas
  under two or more correlated receiver operating characteristic curves: A
  nonparametric approach}, {\emph{Biometrics} {\bfseries 44} (1988) 837}.

\bibitem{6851192}
X.~Sun and W.~Xu, \emph{Fast implementation of delong’s algorithm for
  comparing the areas under correlated receiver operating characteristic
  curves}, \href{https://doi.org/10.1109/LSP.2014.2337313}{\emph{IEEE Signal
  Processing Letters} {\bfseries 21} (2014) 1389}.

\bibitem{ROE2005577}
B.P.~Roe, H.-J.~Yang, J.~Zhu, Y.~Liu, I.~Stancu and G.~McGregor, \emph{Boosted
  decision trees as an alternative to artificial neural networks for particle
  identification},
  \href{https://doi.org/https://doi.org/10.1016/j.nima.2004.12.018}{\emph{Nuclear
  Instruments and Methods in Physics Research Section A: Accelerators,
  Spectrometers, Detectors and Associated Equipment} {\bfseries 543} (2005)
  577}.

\bibitem{wang}
X.~Wang, Y.~Zhao and F.~Pourpanah, \emph{Recent advances in deep learning},
  \href{https://doi.org/10.1007/s13042-020-01096-5}{\emph{International Journal
  of Machine Learning and Cybernetics} {\bfseries 11} (2020) 747}.

\bibitem{Stanev:2021mkr}
D.~Stanev, R.~Riva and M.~Umassi, \emph{{Deep Neural Network as an alternative
  to Boosted Decision Trees for PID}},
  \href{https://arxiv.org/abs/2104.14045}{{\ttfamily 2104.14045}}.

\bibitem{Alvestad:2021sje}
D.~Alvestad, N.~Fomin, J.~Kersten, S.~Maeland and I.~Str\"umke, \emph{{Beyond
  Cuts in Small Signal Scenarios - Enhanced Sneutrino Detectability Using
  Machine Learning}},  \href{https://arxiv.org/abs/2108.03125}{{\ttfamily
  2108.03125}}.

\bibitem{Tannenwald:2020mhq}
B.~Tannenwald, C.~Neu, A.~Li, G.~Buehlmann, A.~Cuddeback, L.~Hatfield et~al.,
  \emph{{Benchmarking Machine Learning Techniques with Di-Higgs Production at
  the LHC}},  \href{https://arxiv.org/abs/2009.06754}{{\ttfamily 2009.06754}}.

\bibitem{Heredge:2021vww}
J.~Heredge, C.~Hill, L.~Hollenberg and M.~Sevior, \emph{{Quantum Support Vector
  Machines for Continuum Suppression in B Meson Decays}},
  \href{https://arxiv.org/abs/2103.12257}{{\ttfamily 2103.12257}}.

\bibitem{Terashi:2020wfi}
K.~Terashi, M.~Kaneda, T.~Kishimoto, M.~Saito, R.~Sawada and J.~Tanaka,
  \emph{{Event Classification with Quantum Machine Learning in High-Energy
  Physics}}, \href{https://doi.org/10.1007/s41781-020-00047-7}{\emph{Comput.
  Softw. Big Sci.} {\bfseries 5} (2021) 2}
  [\href{https://arxiv.org/abs/2002.09935}{{\ttfamily 2002.09935}}].

\bibitem{Bendavid:2017zhk}
J.~Bendavid, \emph{{Efficient Monte Carlo Integration Using Boosted Decision
  Trees and Generative Deep Neural Networks}},
  \href{https://arxiv.org/abs/1707.00028}{{\ttfamily 1707.00028}}.

\bibitem{Strong:2020mge}
G.C.~Strong, \emph{{On the impact of selected modern deep-learning techniques
  to the performance and celerity of classification models in an experimental
  high-energy physics use case}},
  \href{https://arxiv.org/abs/2002.01427}{{\ttfamily 2002.01427}}.

\bibitem{lumin}
G.C.~Strong, \emph{Gilesstrong/lumin: v0.8.0 - mistake not...}, .

\bibitem{8037515}
C.-Y.~Hung, W.-C.~Chen, P.-T.~Lai, C.-H.~Lin and C.-C.~Lee, \emph{Comparing
  deep neural network and other machine learning algorithms for stroke
  prediction in a large-scale population-based electronic medical claims
  database},  in \emph{2017 39th Annual International Conference of the IEEE
  Engineering in Medicine and Biology Society (EMBC)}, pp.~3110--3113, 2017,
  \href{https://doi.org/10.1109/EMBC.2017.8037515}{DOI}.

\bibitem{abdar}
M.~Abdar, N.Y.~Yen and J.C.-S.~Hung, \emph{Improving the diagnosis of liver
  disease using multilayer perceptron neural network and boosted decision
  trees}, \href{https://doi.org/10.1007/s40846-017-0360-z}{\emph{Journal of
  Medical and Biological Engineering} {\bfseries 38} (2018) 953}.

\bibitem{alexei}
A.~Botchkarev, \emph{Evaluating hospital case cost prediction models using
  azure machine learning studio}, {\emph{CoRR} {\bfseries abs/1804.01825}
  (2018) } [\href{https://arxiv.org/abs/1804.01825}{{\ttfamily 1804.01825}}].

\bibitem{chaochao}
C.~Chen, Z.~Liu, J.~Zhou, X.~Li, Y.~Qi, Y.~Jiao et~al., \emph{How much can {A}
  retailer sell? sales forecasting on tmall}, {\emph{CoRR} {\bfseries
  abs/2002.11940} (2020) } [\href{https://arxiv.org/abs/2002.11940}{{\ttfamily
  2002.11940}}].

\bibitem{partin}
A.~Partin, T.~Brettin, Y.A.~Evrard, Y.~Zhu, H.~Yoo, F.~Xia et~al.,
  \emph{Learning curves for drug response prediction in cancer cell lines},
  \href{https://doi.org/10.1186/s12859-021-04163-y}{\emph{BMC Bioinformatics}
  {\bfseries 22} (2021) 252}.

\bibitem{water}
Y.~Chen, W.~Chen, S.C.~Pal, A.~Saha, I.~Chowdhuri, B.~Adeli et~al.,
  \emph{Evaluation efficiency of hybrid deep learning algorithms with neural
  network decision tree and boosting methods for predicting groundwater
  potential},
  \href{https://doi.org/10.1080/10106049.2021.1920635}{\emph{Geocarto
  International} {\bfseries 0} (2021) 1}.

\bibitem{Carleo:2019ptp}
G.~Carleo, I.~Cirac, K.~Cranmer, L.~Daudet, M.~Schuld, N.~Tishby et~al.,
  \emph{{Machine learning and the physical sciences}},
  \href{https://doi.org/10.1103/RevModPhys.91.045002}{\emph{Rev. Mod. Phys.}
  {\bfseries 91} (2019) 045002}
  [\href{https://arxiv.org/abs/1903.10563}{{\ttfamily 1903.10563}}].

\bibitem{Rajkomar2019-vl}
A.~Rajkomar, J.~Dean and I.~Kohane, \emph{Machine learning in medicine},
  {\emph{N. Engl. J. Med.} {\bfseries 380} (2019) 1347}.

\bibitem{keith}
J.A.~Keith, V.~Vassilev-Galindo, B.~Cheng, S.~Chmiela, M.~Gastegger,
  K.-R.~M{\"u}ller et~al., \emph{Combining machine learning and computational
  chemistry for predictive insights into chemical systems},
  \href{https://doi.org/10.1021/acs.chemrev.1c00107}{\emph{Chemical Reviews}
  {\bfseries 121} (2021) 9816}.

\bibitem{QIAN2021100642}
X.~Qian and R.~Yang, \emph{Machine learning for predicting thermal transport
  properties of solids},
  \href{https://doi.org/https://doi.org/10.1016/j.mser.2021.100642}{\emph{Materials
  Science and Engineering: R: Reports} {\bfseries 146} (2021) 100642}.

\bibitem{NASIRI20211137}
S.~Nasiri and M.R.~Khosravani, \emph{Machine learning in predicting mechanical
  behavior of additively manufactured parts},
  \href{https://doi.org/https://doi.org/10.1016/j.jmrt.2021.07.004}{\emph{Journal
  of Materials Research and Technology} {\bfseries 14} (2021) 1137}.

\bibitem{Dudley2018ARO}
J.J.~Dudley and P.O.~Kristensson, \emph{A review of user interface design for
  interactive machine learning}, {\emph{ACM Transactions on Interactive
  Intelligent Systems (TiiS)} {\bfseries 8} (2018) 1 }.

\end{thebibliography}\endgroup

\newpage
\appendix

\section{Description of features}
\label{app:highlow}
In this appendix we show how the options of ``high-level features'', ``all features'' and ``low-level features'' are designed. In Tab.~\ref{tab:featuresapp} there is a tick for each feature that is used for the specific option. The option ``all features'' includes, as the name implies, all of the available features we simulate. In the other cases we tried to use enough features to make the options distinguishable to each other but still being able to separate the different instances of the data. 
%When using high-level features we erase from the all features options all the redundant features that are easily calculated. However when using low-level features we erased all the features that can be calculated using the ones we did select, leaving it in four total features. 
As explained in the main text, we create these options in order to prove how different the models behave when giving redundant information versus how they work when giving just the necessary information.

\begin{table}[H]
\small
\centering
\caption{Features present in each training option of the data. Note the isolation only applies to the $\bmu$ channel.\label{tab:featuresapp}}
\resizebox{\myfigwidth \textwidth}{!}{
\begin{tabular}{llll}
\toprule
     &   high-level features & all features & low-level  features \\ 
     \midrule
$p_{TB}$ &  \checkmark &  \checkmark  &     \\ 
$p_{Tdaug}$   &  \checkmark &  \checkmark &  \\   
$IP_{B}$    &  \checkmark &  \checkmark  &   \\   
$IP_{\rm{daughters}}$   &  \checkmark &  \checkmark  &   \\   
Isolation $\mu$  &  \checkmark &  \checkmark  &  \checkmark \\   
Position of daughters  &   &  \checkmark  &  \checkmark \\   
Position of mothers   &   &  \checkmark  &  \checkmark \\   
Doca  &  \checkmark &  \checkmark  &   \\   
DoF &  \checkmark &  \checkmark  &   \\   
$p_x,p_y,p_z$ daughters  &   &  \checkmark &  \checkmark\\  
\bottomrule
\end{tabular}}
\end{table}

%%%%%%%%%%%%%%%%%%%%%%%%%%%%%%%%%%%%%%%%%%%%%%%%%%%%%%%%%%%%%%%%%%%%%%%%%%%%%%%%%%%%%%%%%%%%%%%%%%%%%%%%
%%%%%%%%%%%%%%%%%%%%%%%%%%%%%%%%%%%%%%%%%%%%%%%%%%%%%%%%%%%%%%%%%%%%%%%%%%%%%%%%%%%%%%%%%%%%%%%%%%%%%%%%
%%%%%%%%%%%%%%%%%%%%%%%%%%%%%%%%%%%%%%%%%%%%%%%%%%%%%%%%%%%%%%%%%%%%%%%%%%%%%%%%%%%%%%%%%%%%%%%%%%%%%%%%
%%%%%%%%%%%%%%%%%%%%%%%%%%%%%%%%%%%%%%%%%%%%%%%%%%%%%%%%%%%%%%%%%%%%%%%%%%%%%%%%%%%%%%%%%%%%%%%%%%%%%%%%
%%%%%%%%%%%%%%%%%%%%%%%%%%%%%%%%%%%%%%%%%%%%%%%%%%%%%%%%%%%%%%%%%%%%%%%%%%%%%%%%%%%%%%%%%%%%%%%%%%%%%%%%

\section{Hyperparameters for the classifiers.}
\label{sec:hyper}
In this appendix we show all the hyperparameters needed by the libraries that we selected after an exhaustive grid search. For each decay, training option and tool the results are shown in the following tables, to allow the reader to replicate our results and estimate for which values these algorithms work better for the correspondent problem and decay.

\begin{table}[H]
\small\caption{Neural Network parameters for PyTorch for the $\bmu$ decay.  \label{tab:pbmumuPytorch}}
    \centering
        \resizebox{\myfigwidth \textwidth}{!}{
    \begin{tabular}{llllll}
      \toprule
     & A. function  & Hidden layers & Learning rate & Momentum & Epochs \\
\midrule
        low-level features - low stats & ReLU  &  9   &   0.0089 & 0.75 & 100    \\   
 low-level features - high stats   &  ReLU   &  12   &  0.0093 & 0.8 & 180  \\   
 high-level features - low stats  &  ReLU    &  11   &  0.009  & 0.9 & 80\\   
 high-level features - high stats &  Tanh    &  9    & 0.0086 & 0.8 & 120\\
   
 all features - low stats  &  ReLU    &  14    &  0.0075 & 0.85 & 110 \\   
 all features - high stats &  Tanh   &  12  &  0.0092 & 0.85 & 170 \\  
 \bottomrule
    \end{tabular}}
\end{table}

\begin{table}[H]
\small
 \caption{Neural Network parameters for Keras for the $\bmu$ decay.  \label{tab:pbmumuKeras}}
    \centering
    \resizebox{\myfigwidth \textwidth}{!}{
    \begin{tabular}{lllllll}
      \toprule
     & A. function  & Hidden layers & Learning rate & Momentum & Epochs \\
      \midrule
        low-level features - low stats & ReLU  &  8   &   0.008 & 0.8 & 110    \\   
 low-level features - high stats   &  Tanh   &  12   &  0.0085 & 0.85 & 180  \\   
 high-level features - low stats  &  ReLU    &  11   &  0.0087  & 0.8 & 100\\   
 high-level features - high stats &  Tanh    &  9    & 0.0082 & 0.8 & 130\\
   
 all features - low stats  &  ReLU    &  14    &  0.0089 & 0.85 & 100 \\   
 all features - high stats &  ReLU   &  12  &  0.0082 & 0.9 & 190 \\  
 \bottomrule
    \end{tabular}}
   \end{table}

\begin{table}[H]
\small
   \caption{Neural Network parameters for PyTorch for the $\bpipi$ decay.  \label{tab:pbpiPytorch}}
    \centering
    \resizebox{\myfigwidth \textwidth}{!}{
    \begin{tabular}{llllll}
      \toprule
     & A. function  & Hidden layers & Learning rate & Momentum & Epochs \\
      \midrule
        low-level features - low stats & Tanh  &  14   &   0.0089 & 0.85 & 100    \\   
 low-level features - high stats   &  ReLU   &  10   &  0.0091 & 0.8 & 140  \\   
 high-level features - low stats  &  ReLU    &  10   &  0.0087  & 0.85 & 100\\   
 high-level features - high stats &  Tanh    &  12    & 0.0088 & 0.8 & 120\\
   
 all features - low stats  &  ReLU    &  10    &  0.0085 & 0.95 & 100 \\   
 all features - high stats &  Tanh   &  11  &  0.0089 & 0.85 & 130 \\  
 \bottomrule
    \end{tabular}}
\end{table}

\begin{table}[H]
\small
\caption{Neural Network parameters for Keras for the $\bpipi$ decay.  \label{tab:pbpiKeras}}
    \centering
     \resizebox{\myfigwidth \textwidth}{!}{
    \begin{tabular}{llllll}
      \toprule
     & A. function  & Hidden layers & Learning rate & Momentum & Epochs \\
      \midrule
        low-level features - low stats & ReLU  &  10   &   0.0079 & 0.85 & 100    \\   
 low-level features - high stats   &  Tanh   &  14   &  0.0081 & 0.85 & 130  \\   
 high-level features - low stats  &  ReLU    &  10   &  0.0091  & 0.9 & 110\\   
 high-level features - high stats &  Tanh    &  16   & 0.0093 & 0.85 & 140\\
   
 all features - low stats  &  ReLU    &  11    &  0.009 & 0.95 & 120 \\   
 all features - high stats &  ReLU   &  12  &  0.0086 & 0.85 & 120 \\  
 \bottomrule
    \end{tabular}}
    \end{table}

\begin{table}[H]
\small
\caption{Neural Network parameters for PyTorch for the $\btripi$ decay.  \label{tab:pbtpiPytorch}}
    \centering
    \resizebox{\myfigwidth \textwidth}{!}{
    \begin{tabular}{llllll}
      \toprule
     & A. function  & Hidden layers & Learning rate & Momentum & Epochs \\
      \midrule
        low-level features - low stats & Tanh  &  11   &   0.009 & 0.9 & 110    \\   
 low-level features - high stats   &  ReLU   &  13   &  0.009 & 0.85 & 100  \\   
 high-level features - low stats  &  ReLU    &  12   &  0.0085  & 0.85 & 110\\   
 high-level features - high stats &  Tanh    &  12    & 0.0087 & 0.75 & 130\\
   
 all features - low stats  &  ReLU    &  12    &  0.0089 & 0.9 & 110 \\   
 all features - high stats &  Tanh   &  16  &  0.0086 & 0.8 & 130 \\ 
 \bottomrule
    \end{tabular}}
\end{table}

\begin{table}[H]
\small
\caption{Neural Network parameters for Keras for the $\btripi$ decay.  \label{tab:pbtpiKeras}}
    \centering
    \resizebox{\myfigwidth \textwidth}{!}{
    \begin{tabular}{llllll}
      \toprule
     & A. function  & Hidden layers & Learning rate & Momentum & Epochs \\
      \midrule
        low-level features - low stats & ReLU  &  9   &   0.0089 & 0.8 & 100    \\   
 low-level features - high stats   &  Tanh   &  13  &  0.008 & 0.8 & 120  \\   
 high-level features - low stats  &  ReLU    &  10   &  0.009  & 0.85 & 130\\   
 high-level features - high stats &  Tanh    &  12    & 0.0087 & 0.85 & 150\\
 all features - low stats  &  ReLU    &  11    &  0.0091 & 0.9 & 100 \\   
 all features - high stats &  ReLU   &  15  &  0.0085 & 0.8 & 140 \\  
 \bottomrule
    \end{tabular}}
\end{table}

\begin{table}[H]
\small
    \caption{Neural Network parameters for PyTorch for the $\bcpi$ decay.  \label{tab:pbcpiPytorch}}
    \centering
     \resizebox{\myfigwidth \textwidth}{!}{
    \begin{tabular}{llllll}
      \toprule
     & A. function  & Hidden layers & Learning rate & Momentum & Epochs \\
      \midrule
        low-level features - low stats & Tanh  &  11   &   0.009 & 0.81 & 100    \\   
 low-level features - high stats   &  ReLU   &  13   &  0.0092 & 0.85 & 140  \\   
 high-level features - low stats  &  ReLU    &  10   &  0.0089  & 0.85 & 110\\   
 high-level features - high stats &  Tanh    &  10    & 0.0078 & 0.8 & 150\\
   
 all features - low stats  &  ReLU    &  10    &  0.0075 & 0.9 & 130 \\   
 all features - high stats &  Tanh   &  14  &  0.0079 & 0.85 & 160 \\  
 \bottomrule
    \end{tabular}}
\end{table}

\begin{table}[H]
\small
 \caption{Neural Network parameters for Keras for the $\bcpi$ decay.  \label{tab:pbcpiKeras}}
    \centering
      \resizebox{\myfigwidth \textwidth}{!}{
    \begin{tabular}{llllll}
      \toprule
     & A. function  & Hidden layers & Learning rate & Momentum & Epochs \\
      \midrule
        low-level features - low stats & ReLU  &  10   &   0.008 & 0.8 & 100    \\   
 low-level features - high stats   &  Tanh   &  16   &  0.0082 & 0.8 & 120  \\   
 high-level features - low stats  &  ReLU    &  9   &  0.0081  & 0.7 & 100\\   
 high-level features - high stats &  Tanh    &  12    & 0.0095 & 0.85 & 160\\
   
 all features - low stats  &  ReLU    &  10    &  0.009 & 0.9 & 110 \\   
 all features - high stats &  ReLU   &  9  &  0.0086 & 0.9 & 140 \\  \bottomrule
    \end{tabular}}
   
\end{table}

\begin{table}[H]
\small
\caption{BDT with AdaBoost parameters for Sklearn for the $\bmu$ decay.  \label{tab:pbmuSklearn}}
    \centering
     \resizebox{\myfigwidth \textwidth}{!}{
     \begin{tabular}{llll}
      \toprule
     &  Base estimator  & Nº estimators & Learning rate   \\
      \midrule
        low-level features - low stats & DecisionTreeClassifier   &  120   &   0.8    \\   
 low-level features - high stats   &  DecisionTreeClassifier    &  100   &  0.8  \\   
 high-level features - low stats  &  DecisionTreeClassifier     &  150   &  0.7 \\   
 high-level features - high stats &  DecisionTreeClassifier     &  170    & 0.9 \\
   
 all features - low stats  &  DecisionTreeClassifier     &  160    &  0.9  \\   
 all features - high stats &  DecisionTreeClassifier    &  200  &  0.8  \\ 
 \bottomrule
    \end{tabular}}
\end{table}

\begin{table}[H]
\small
\caption{BDT with AdaBoost parameters for Sklearn for the $\bpipi$ decay.  \label{tab:pbpiSklearn}}
\centering
    \resizebox{\myfigwidth \textwidth}{!}{
    \begin{tabular}{llll}
      \toprule
     &  Base estimator  & Nº estimators & Learning rate   \\
      \midrule
        low-level features - low stats & DecisionTreeClassifier   &  100   &   0.8    \\   
 low-level features - high stats   &  DecisionTreeClassifier    &  130   &  0.8  \\   
 high-level features - low stats  &  DecisionTreeClassifier     &  150   &  0.8 \\   
 high-level features - high stats &  DecisionTreeClassifier     &  200    & 0.9 \\
   
 all features - low stats  &  DecisionTreeClassifier     &  170    &  0.9  \\   
 all features - high stats &  DecisionTreeClassifier    &  250  &  0.9  \\  
 \bottomrule
    \end{tabular}}
    \end{table}

\begin{table}[H]
\small
\caption{BDT with AdaBoost parameters for Sklearn for the $\btripi$ decay.  \label{tab:pbtpiSklearn}}
    \centering
    \resizebox{\myfigwidth \textwidth}{!}{
    \begin{tabular}{llll}
      \toprule
     &  Base estimator  & Nº estimators & Learning rate   \\
      \midrule
        low-level features - low stats & DecisionTreeClassifier   &  110   &   0.75    \\   
 low-level features - high stats   &  DecisionTreeClassifier    &  170   &  0.9  \\   
 high-level features - low stats  &  DecisionTreeClassifier     &  190   &  0.75 \\   
 high-level features - high stats &  DecisionTreeClassifier     &  230    & 0.9 \\
   
 all features - low stats  &  DecisionTreeClassifier     &  220    &  0.85  \\   
 all features - high stats &  DecisionTreeClassifier    &  240  &  0.9  \\
 \bottomrule
    \end{tabular}}
\end{table}

\begin{table}[H]
\small
 \caption{BDT with AdaBoost parameters for Sklearn for the $\bcpi$ decay.  \label{tab:pbcpiSklearn}}
    \centering
    \resizebox{\myfigwidth \textwidth}{!}{
    \begin{tabular}{llll}
      \toprule
     &  Base estimator  & Nº estimators & Learning rate   \\
      \midrule
        low-level features - low stats & DecisionTreeClassifier   &  120   &   0.85    \\   
 low-level features - high stats   &  DecisionTreeClassifier    &  120   &  0.85  \\   
 high-level features - low stats  &  DecisionTreeClassifier     &  150   &  0.8 \\   
 high-level features - high stats &  DecisionTreeClassifier     &  190    & 0.9 \\
   
 all features - low stats  &  DecisionTreeClassifier     &  160    &  0.95  \\   
 all features - high stats &  DecisionTreeClassifier    &  230  &  0.9  \\  
 \bottomrule
    \end{tabular}}
\end{table}

\section{Training time for the classifiers in a personal laptop for the \texorpdfstring{$\bmu$}{Bmumu} decay}
\label{sec:time}
In order to obtain another metric of how good an algorithm is we present an example of how long it takes an algorithm to train with the samples we selected. We present all the different analyses for the decay $\bmu$. Comparable times are obtained for other decays.
\begin{table}[H]
\small
\centering
\caption{Training time for each tool for the decay $\bmu$ \label{tab:time}}
\resizebox{\myfigwidth \textwidth}{!}{
\begin{tabular}{llll}
\toprule
& \textbf{PyTorch} & \textbf{Keras} & \textbf{Sklearn} \\ 
\midrule
high-level features - high stats  & 5 min 56 s &  5 min 46 s& 5 min 43 s\\   
high-level features - low stats & 4 min 35 s &4 min 43 s   & 4 min 30s     \\   
low-level features - low stats & 4 min 05 s &4 min 13 s   & 4 min 11s     \\   
low-level features - high stats & 5 min 34 s  & 5 min 51 s &    6 min 36 s\\   
all features - high stats & 6 min 06 s & 6 min 11 s  & 5 min 58 s   \\   
all features - low stats &  6 min 46 s & 6 min 03 s   & 5 min 20s    \\   
\bottomrule
\end{tabular}}

\end{table}

\end{document}